\documentclass[prb,aps,nobalancelastpage,superscriptaddress,twocolumn,longbibliography]{revtex4-2}

\usepackage[utf8]{inputenc}
\usepackage[american,british]{babel}
\usepackage[T1]{fontenc}
\usepackage[pdftex]{graphicx}  
\usepackage{graphicx, xcolor}
\usepackage{dcolumn}
\usepackage{physics}
\usepackage{braket}
\usepackage{bm}
\usepackage{amsmath,amsthm,amssymb}
\usepackage{color}
\usepackage{verbatim}
\usepackage[normalem]{ulem}

\usepackage{hyperref}
\hypersetup{
 colorlinks=true,
 linkcolor=blue,
 anchorcolor = blue,
 citecolor = blue,
 filecolor = blue,
 urlcolor = blue 
}

\def \be {\begin{equation}} 
\def \ee {\end{equation}} 
\def \l {\left(} 
\def \r {\right)} 
\def \la {\langle} 
\def \ra {\rangle}

\date{}

\def\lptms{Universit\'e Paris-Saclay, CNRS, LPTMS, 91405, Orsay, France.}

\def\iuf{Institut Universitaire de France, 75005, Paris, France.}

\begin{document}

\title{Universal properties of the many-body Lanczos algorithm at finite size}
\date{\today}

\author{Luca Capizzi}
\affiliation{\lptms}
\author{Leonardo Mazza}
\affiliation{\lptms}
\affiliation{\iuf}
\author{Sara Murciano}
\affiliation{\lptms}

\begin{abstract}
We study the universal properties of the Lanczos algorithm applied to finite-size many-body quantum systems.
Focusing on autocorrelation functions of local operators and on their infinite-time behaviour at finite size,
we conjecture that in the large $n$ limit, the ratios between consecutive Lanczos coefficients should have specific scalings with the size of the lattice that we make precise and that depend on the hydrodynamic tail of the autocorrelation function.
The scaling associated with strong or approximate zero-modes is also discussed.
We support our conjecture with a numerical study of different models.
\end{abstract}

\maketitle 

\section{Introduction}

Understanding many-body quantum dynamics is an outstanding challenge: predicting whether and how a setup will eventually thermalize 
is a problem of exceptional theoretical interest
with ample practical applications in the context of synthetic quantum matter.
So far, several theoretical frameworks can be used to address the late-time behavior of a high-temperature quantum many-body system: from the eigenstate thermalization hypothesis \cite{Deutsch-91, Srednicki-99} to the conventional \cite{km-63,Spohn-12} and generalized hydrodynamics \cite{Castro-16, Bertini-16}; from numerical techniques based on entanglement~\cite{Cirac_2021} or dissipation~\cite{Rakovszky_2022} to the Lanczos algorithm (LA) \cite{Recursion-1994,Nandy-2025}.

In the last years, the LA has experienced a revival of interest due to the conjecture that the Lanczos coefficients satisfy a universal behaviour associated to the chaotic or integrable behaviour of the model~\cite{Parker-19}.
Among the other applications, the LA is a useful tool also for understanding the dynamics of the edge modes of different spin chains~\cite{Yates2020PRL,yam-20,yates2021PRB,Yeh2023,tmr-25}, to study Floquet systems~\cite{yeh2024PRB,yeh2025,yates2022comm,Suchsland2025,Kolganov2025}, to probe state and operator complexity in holography~\cite{Rabinovici2023,Balasubramanian2024,Caputa2024,Miyaji2025,Caputa2025} and many-body quantum systems~\cite{Avdoshkin2022,Dymarsky2021,Bhattacharjee2022operator,Balasubramanian2022dnj,Balasubramanian2024ghv,Erdmenger2023,Rabinovici2022,Caputa2025gro}.
The possibility of inferring emerging hydrodynamic effects~\cite{Bhattacharjee_2022,Cao2021,qi2023,Nandy-2025,Uskov2024,wang2024,yi2024,Loizeau2025quantum,loizeau2025buca,Fullgraf2025,Gamayun2025hvu,Shirokov2025, plv-25, lkmv-25} 
has been also discussed. Yet, these approaches suffer from the fact that the Lanczos coefficients that can be computed numerically are soon affected by finite-size effects, whereas the aforementioned studies focused entirely on infinite-system behaviors: this puts a strong constraint on the possibility of utilizing these conjectures in practical numerical simulations.

In this work, we show that the Lanczos coefficients display universal properties even when studied at finite size. 
We extend the \textit{Universal Operator Growth Hypothesis} \cite{Parker-19}, expected to be valid for generic many-body systems in infinite lattices; we find that the Lanczos coefficients follow general \textit{finite-size} patterns that are largely independent from the microscopic details of the interactions. We obtain this result by focusing on the asymptotic infinite-time value of the autocorrelation function of a local observable at finite size and by discussing its relation with the LA. We present a threefold conjecture that covers several situations. We support our analysis and our claim of universality with numerical simulations on several lattice models with qualitatively different properties.
This work opens the path to the use of the LA in finite systems to extract information on the universal late-time properties of quantum many-body systems.

We structure the manuscript as follows: in Sec.~\ref{sec:Prelim}, we introduce the theoretical framework and summarize the Lanczos algorithm, focusing on its application to autocorrelation functions and the infinite-time behavior of local observables. In Sec.~\ref{sec:conjecture}, we present our central conjecture, which connects the late-time plateau of autocorrelation functions to the scaling of Lanczos coefficients, and discuss its implications for hydrodynamic behavior, vanishing plateaus, and strong zero modes. In Sec.~\ref{sec:Anal_arg}, we provide analytical arguments to support our conjecture, while Sec.~\ref{sec:Numerics} is dedicated to numerical benchmarks across various models, including long-range and higher-dimensional systems. We discuss conclusions and outlook in Sec.~\ref{sec:Conclusions} and address technical details of the numerics in Appendix \ref{app:numerics}.

\section{Preliminaries}\label{sec:Prelim}
We begin by summarizing the Lanczos algorithm~\cite{Recursion-1994, Nandy-2025}.
Consider a quantum many-body system with Hamiltonian $H$ defined on a finite lattice of size $L^d$, with $d$ the dimensionality of the setup and $L$ its linear dimension.
We focus on the infinite-temperature state, $\la \mathcal{O}\ra := \text{Tr}(\mathcal{O})/\text{Tr}(1)$; generalizations to finite temperature are possible but will not be considered here.
We study the autocorrelation function of a generic local observable $\mathcal{O}$:
\begin{equation}\label{eq:Autocorr_L}
 C_L(t) := \la \mathcal{O}(t)\mathcal{O}\ra-\la \mathcal{O}(t)\ra\la\mathcal{O}\ra,
\end{equation}
where the operator $\mathcal{O}(t)$ is time-evolved in the Heisenberg picture and obeys $\partial_t \mathcal{O}(t) = i [H , \mathcal{O}(t)]$. 
In Eq.~\eqref{eq:Autocorr_L} we explicitly state the dependence on $L$ because it plays a key role in the forthcoming discussions.

The key idea of the LA is to represent the superoperator $\mathcal L [\cdot] := [H,\cdot]$ as a tridiagonal matrix
\begin{equation}
 \mathcal L =
 \begin{pmatrix}
	0 & b_1 & 0 & 0 & \cdots\\
	b_1 & 0 & b_2 & 0 & \cdots\\
	0 & b_2 & 0 & b_3 & \cdots\\[-0.3em]
	0 & 0 & b_3 & 0 & \ddots\\[-0.2em]
	\vdots & \vdots & \vdots & \ddots & \ddots
	\end{pmatrix},
 \label{Eq:MatRep}
\end{equation}
with $b_n>0$ known as \textit{Lanczos coefficients}.
Specifically, an orthonormal basis of operators (with respect to the product $(\mathcal{O},\mathcal{O}') := \la\mathcal{O}^\dagger\mathcal{O}'\ra$) is constructed from the Gram–Schmidt orthogonalization algorithm applied to the nested commutators between $H$ and $\mathcal{O}$ (see Ref.~\cite{Parker-19}).
We will focus on observables with a vanishing expectation value ($\langle \mathcal{O} \rangle = 0$) and unit norm ($\langle \mathcal{O}^\dagger \mathcal{O} \rangle = 1$).

\begin{figure}[t]
\includegraphics[width=\columnwidth]{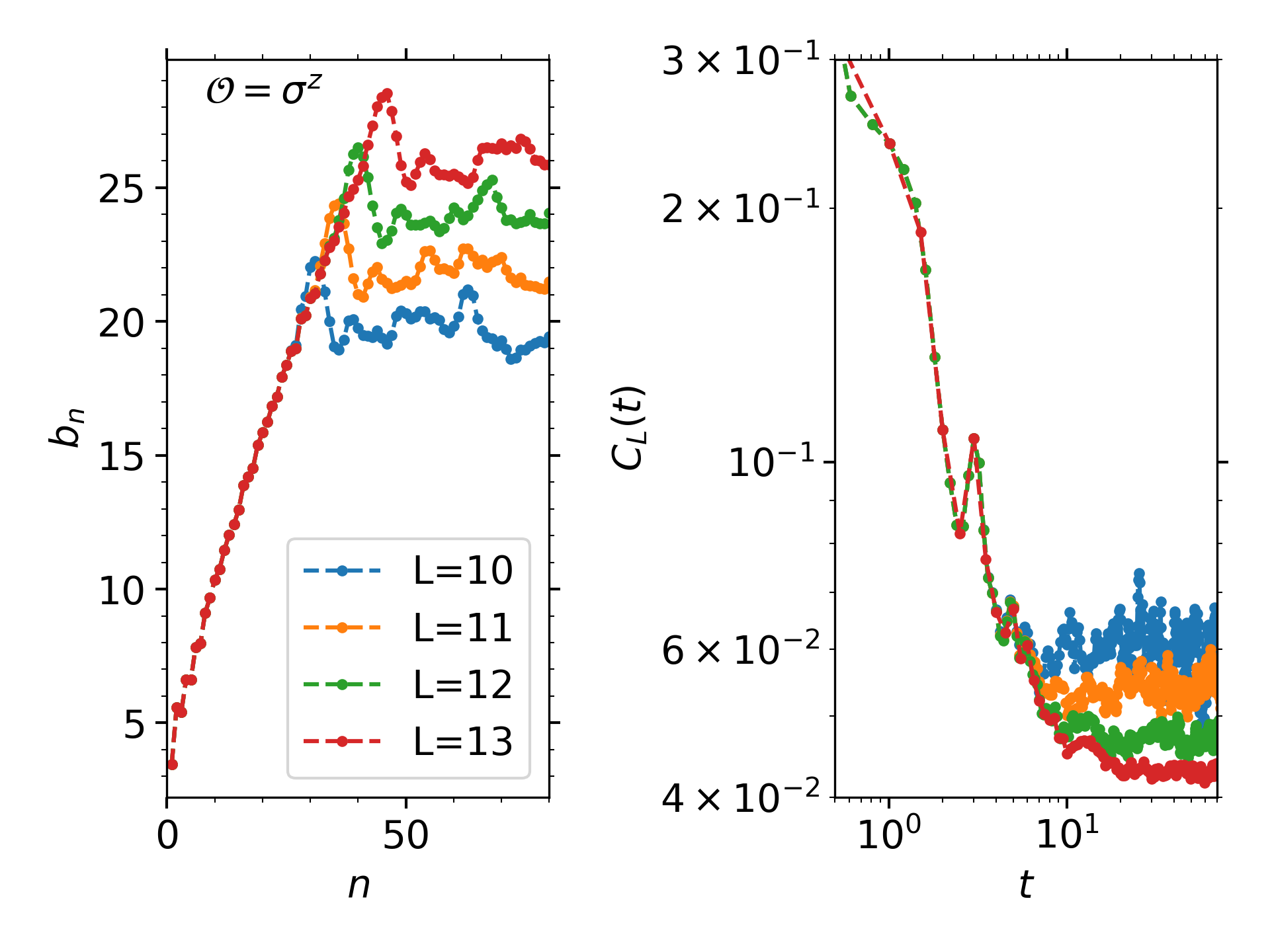}
 \caption{Left: Lanczos coefficients computed for model~\eqref{Eq:TFIM} for several sizes, from $L=10$ to $L=13$, with periodic boundary conditions. The observable is $\sigma^z_1$ and the parameters are $[J,h_x,h_z] = [1,1,1.5]$. The plot clearly shows the presence of an infinite-size behaviour for $n<n^*$ followed by a fluctuating plateau region affected by finite-size effects.
 Right: Corresponding correlation functions $C_L(t)$ as a function of time $t$; a size-dependent plateau is observed at late times.
 }
 \label{Fig:1:Example}
\end{figure}

  As an example, we consider the following non-integrable one-dimensional Ising model with transverse and longitudinal magnetic fields:
\begin{equation}
H = \sum_j \left( J  \sigma_j^x  \sigma_{j+1}^x
 + h_x  \sigma_j^x + h_z  \sigma_j^z \right),
 \label{Eq:TFIM}
\end{equation}
where $\sigma^{\alpha}_j$, $\alpha=x,y,z$, are the Pauli operators at the $j$-th site.
In the left panel of Fig.~\ref{Fig:1:Example} we show the Lanczos coefficients 
for the operator $\sigma^z_1$ computed for finite chains up to $L=13$ with periodic boundary conditions.
The plot displays the initial universal behavior independent of $L$, which persists in the thermodynamic limit, and the subsequent plateau, which is a finite-size effect. 
The crossover can be intuitively understood because the application of subsequent commutators of $H$ progressively increases the support of $\mathcal{O}$ and indeed takes place at $n^*$ of the order of $L$. 
Similar properties are also found for the other operators that we will consider, $\sigma^y_1$ and $\sigma^z_1\sigma^z_2$. 
As a comparison, we plot the corresponding correlation functions $C_L(t)$ in the right panel of Fig.~\ref{Fig:1:Example}: these (approximately) coincide for $t$ sufficiently small, and then show a plateau depending on the system size $L$.

Whereas so far many of the studies have focused on the universal properties of the Lanczos coefficients for $n < n^*$, current numerical techniques can reliably compute only a few tens of them, so that large $n$ scalings are numerically difficult to address.
The goal of this work is to identify the universal scalings that characterize the coefficients for $n>n^*$.
Specifically, we want to understand the relation between the plateau of the autocorrelation functions and the corresponding Lanczos coefficients.

\section{The conjecture}\label{sec:conjecture}
We focus on the late-time limit of $C_{ L}(\infty)=\lim_{t \to \infty }C_L(t)$, and in particular on its behavior as a function of $L$.
We begin by stating a formula, derived in Sec.~\ref{sec:proof} (see also Ref.~\cite{Gamayun2025hvu}), that relates exactly $C_{L}(\infty)$ to the Lanczos coefficients $b_n$:
\begin{equation}
 C_{L} (\infty) = \frac{1}{1+
 \sum_{n=1}^{\infty}
 \prod_{m=1}^n
 \left( \frac{b_{2m-1}}{b_{2m}} \right)^2}.
 \label{Eq:InfiniteTimeFormula}
\end{equation}
This formula is amenable to a simple interpretation: the Lanczos Hamiltonian~\eqref{Eq:MatRep}
has a zero-energy eigenvector localised at the boundary of the fictitious Lanczos chain;
the denominator of~\eqref{Eq:InfiniteTimeFormula} is simply its squared norm. 
We can conclude that (i) $C_{L} (\infty)$ is different from zero when such a mode exists and is normalizable, and that (ii) $C_{L} (\infty) = 0$ when it does not exist, namely its norm blows to infinity.

The convergence of the infinite sum at the denominator of~\eqref{Eq:InfiniteTimeFormula} ultimately depends on the large-$n$ behavior of the (squared) ratios between odd and even Lanczos coefficients. We introduce the notation
\begin{equation}
   e^{- \Gamma_n (L)} := \left(
  \frac{b_{n}}{b_{n+1}}
  \right)^2 \quad \text{with } n \text{ odd};
  \label{eq:rate}
\end{equation}
here, $\Gamma_n(L)$ plays the role of a rate since, whenever it is constant in $n$, the cumulative products $ \prod_n (b_{n}/b_{n+1})^2$ decay or blow exponentially. In general, its behavior rules the (finite-size) properties of the plateau $C_L(\infty)$, and we formulate the following threefold conjecture for various relevant physical conditions.

\textit{Conjecture 1: Hydrodynamic behaviour.}

In generic closed many-body systems the energy is conserved and the associated hydrodynamic transport always takes place. 
As a consequence, the autocorrelators of a local operator $\mathcal{O}$ decay generically algebrically in time as $\sim t^{-\nu}$ reaching eventually a plateau of order $C_L(\infty)\sim L^{-md}$; the integer number $m$ is identified by the first power of the Hamiltonian $H$ that overlaps with $\mathcal{O}$, satisfying $\la \mathcal{O}H^{m}\ra_c \neq 0$, where $\la \dots\ra_c$ denotes the connected cumulant. 
This mechanism was elucidated in Ref.~\cite{cwxmp-25}, where a conjecture, named \textit{relaxation-overlap inequality}, has been proposed: it predicts $\nu \leq dm/z$, with $z$ the dynamical critical exponent.

The behavior of the Lanczos coefficients for $n<n^*$ in models displaying a hydrodynamic behavior has been deeply investigated.
Generically, these models display
at leading order $b_n\sim n$ (for $d=1$ the scaling is $b_n \sim n/ \ln{n}$~\cite{Parker-19}); 
the subleading staggering proportional to $(-1)^n$ has been recently elucidated in Refs.~\cite{plv-25,lkmv-25}, and it is directly associated with the algebraic decay $\sim t^{-\nu}$. 

Our focus is instead on $n > n^*$.
In this scenario, where the model is chaotic (non-integrable) and the finite-size plateau of $C_{L}(\infty)$ is algebraically small in the system size, we propose the following conjecture
\be\label{eq:conj_hydro}
\Gamma_n(L) \sim L^{-(md+1)}, \quad n>n^{*}.
\ee
In particular, $\Gamma_n(L)$ should be independent of $n$ and positive, while consequently the ratio between odd and even Lanczos coefficients should be smaller than $1$. 
This prediction is expected to be true \textit{on average}, meaning that fluctuations around this leading trend are possible and indeed found in the numerical data.
Technical details relating our conjecture~\eqref{eq:conj_hydro} to the predictions for $n<n^*$ are discussed in Sec.~\ref{sub:technical}.

\textit{Conjecture 2: Vanishing late-time plateau.}

Even if transport is present, local operators do not show a hydrodynamic behavior when they do not overlap with conserved charges. In these situations, $C_L(\infty)$ vanishes or is exponentially small in system size, and we conjecture
\be\label{eq:Conj_van}
\Gamma_n(L) \leq \frac{\alpha}{n}, \quad 
\text{with}
\quad
0<\alpha<2.
\ee
Specifically, it is possible either that $\Gamma_n$ is positive but converges to $0$ sufficiently fast, or that $\Gamma_n$ becomes negative at large $n$; in both cases, the denominator of \eqref{Eq:InfiniteTimeFormula} diverges and the plateau vanishes.

\textit{Conjecture 3: Strong zero modes.}

Recently, situations where $C_L(\infty)$ is nonzero and independent of system size have attracted considerable attention. 
This is what happens in the presence of conserved local operators, known as \textit{strong zero modes}~\cite{yam-20, tmr-25}, where 
the system retains memory of the initial state even in the thermodynamic limit (see e.g.~\cite{SSH-79,SSH-80}).
For boundary fermionic operators, the LA produces a sequence of
$b_n$ that oscillate perfectly between two values and terminates after $n \sim L$~\cite{Parker-19}. 

Additionally, in interacting systems researchers have found \textit{approximate} zero modes, 
characterized by a long-lived plateau of the autocorrelation function whose value does not depend on $L$ and that drops to zero after a given long time~\cite{yam-20, tmr-25}. 
There, for $n<n^* \sim L$, the Lanczos coefficients that have been numerically analyzed scale as $b_n \sim  n + (-1)^n c$, with $c>0$. For $n>n^*$ finite-size effects, described in the next section, kick in.

In situations with zero modes, we conjecture
\be\label{eq:conj_zero_mode}
\Gamma_n \geq \frac{\alpha}{n}, \quad 
\text{with} \quad
\alpha>2
\ee
asymptotically in $n$. 
This inequality guarantees mathematically the existence of a strong zero mode: when the zero modes are approximate, violations are expected.

\subsection{Derivation of Eq.~\eqref{Eq:InfiniteTimeFormula} for $C_L(\infty)$}\label{sec:proof}

First, we briefly summarize the relation between the spectrum of $\mathcal{L}$ and the autocorrelation function $C_{L}(t)$: this is standard material, and we refer the reader to Ref.~\cite{Parker-19} for details. Using the Lanczos basis, the first basis element $\ket{1}$ corresponds to the seed operator $\mathcal{O}$ and the autocorrelation function can be written as
\be\label{eq:Lbasis}
C_{L}(t) = \bra{1}e^{i\mathcal{L}t}\ket{1}.\ee
In particular, its Fourier transform coincides with the projection of the resolvent operator of $\mathcal{L}$ onto $\ket{1}$:
\be
C_L(t) = \int^{\infty}_{-\infty} d\omega \ \Phi(\omega)e^{i\omega t},
\ee
with
\be\label{eq:spec_fun}
\Phi(\omega) := \frac{1}{\pi} \text{Im} \bra{1}\frac{1}{\omega-i0^+-\mathcal{L}}\ket{1}
\ee
playing the role of a spectral density. In particular, whenever $\Phi(\omega)$ as a $\delta$-singularity at $\omega=0$, then the time-average of $C_L(t)$ is different from zero and
\be
\begin{split}
C_L(\infty) := \lim_{T\rightarrow \infty} \frac{1}{T}\int^{T/2}_{-T/2}dt \ C_L(t),\\
\Phi(\omega) = C_L(\infty)\delta(\omega) + \dots,
\end{split}
\ee
hold. Given that the singularity above arises whenever $\omega = 0$ belongs to the discrete spectrum of $\mathcal{L}$ (i.e., $\mathcal{L}$ possesses a normalizable zero-mode), studying the necessary conditions for its presence is crucial.

Motivated by that, we try to solve $\mathcal{L}\ket{\phi} = 0$, with $\ket{\phi}$ a column vector with components $\{\phi_n\}_{n>0}$ and $\mathcal{L}$ the matrix in Eq.~\eqref{Eq:MatRep}. This is performed iteratively, since the matrix $\mathcal{L}$ is tridiagonal, and the condition is equivalent to
\be
\begin{cases}
b_1 \phi_2 = 0,\\
b_n \phi_n + b_{n+1} \phi_{n+2} = 0 \quad n\geq 1.
\end{cases}
\ee
As a result, $\phi_n$ vanishes for even $n$, while it can be written efficiently in terms of $\phi_{n-2}$ for $n$ odd as $\phi_{n} = - (b_{n-2}/b_{n-1})\phi_{n-2}$. Putting everything together, and requiring that the state $\ket{\phi}$ is properly normalized, that is $\sum_{n} |\phi_n|^2 = 1$, one finds
\be
|\phi_1|^2 = \frac{1}{1+\left(\frac{b_1}{b_2}\right)^2 +\left(\frac{b_1 b_3}{b_2 b_4}\right)^2 + \dots}.
\ee
By studying the small $\omega$ behavior of $\Phi(\omega)$, we find
\be
\Phi(\omega) = \frac{|\la 1\mid\phi\ra |^2}{\pi} \text{Im}\frac{1}{\omega-i0^+}+\dots = |\phi_1|^2 \delta(\omega)+\dots,
\ee
which allows us to identify $C_{L}(\infty) = |\phi_1|^2$ and to obtain Eq.~\eqref{Eq:InfiniteTimeFormula}.

Lastly, we note a technical but important point. So far, we have implicitly treated $\mathcal{L}$ as if it were a semi-infinite matrix, which potentially can have both discrete and real spectrum. On the other hand, whenever $L$ is finite, that is the relevant case of our discussion, the Hilbert space of operators is finite, although exponentially large in the system size, and the spectrum is always discrete. As a consequence, the Lanczos matrix $\mathcal{L}$ is always finite and the sequence of ${b_n}$ always ends at $n \sim e^{O(L)}$. Nonetheless, the information coming from the last Lanczos coefficients only affects the dynamics of the model at time scales that are exponentially large in the system sizes: these times, where revivals occur due to Poincaré recurrences, are physically irrelevant. In contrast, finite-size effects at order $t \sim L^z$, associated with transport, with dynamical exponent $z$, are much more accessible; therefore, in practice, we can regard $\mathcal{L}$ as if it were a semi-infinite matrix as long as these time-scales are concerned.

\section{Analytical arguments supporting the conjecture}\label{sec:Anal_arg}

We now support the conjecture with some analytical arguments.
We begin by discussing in general the relation between the asymptotic behavior of the ratios defined in Eq.~\eqref{eq:rate} and the finite size plateau in order to identify the conditions to ensure $C_L(\infty)\neq 0$.
For later convenience, we focus on the following possible asymptotics for large $n$
\be\label{eq:Gamma_family}
\Gamma_n \simeq \frac{\alpha}{n^\beta}, 
\quad
\text{with} \quad
\beta \geq 0
\quad \text{for} \quad n \gg n^*
;
\ee
in other words, we focus on those cases where either $\Gamma_n$ decays to zero algebraically in $n$ or converges to a (positive or negative) constant. By studying the behaviour of the cumulative products (details are reported in Sec.~\ref{sub:technical}), we find
\be
\prod^{n}_{n' \text{ odd}}e^{-\Gamma_{n'}} \sim \begin{cases}
e^{-\alpha n^{1-\beta}}, \quad & \text{for } 0\leq\beta<1;\\ 
n^{-\alpha/2}, \quad & \text{for }\beta=1;\\
O(1), \quad & \text{for }\beta>1.
\end{cases}
\label{eq:cum_prod}
\ee
Thus, the series
$
\sum^{\infty}_{n \text{ odd}} \prod^{n}_{n' \text{ odd}}e^{-\Gamma_{n'}} 
$
converges and consequently the autocorrelator has an infinite-time limit different from zero in the two following cases:
\begin{equation}\label{Eq:Table:Conv}
C_L(\infty) \neq 0 \; \Leftrightarrow \;
    \begin{cases}
    0 \leq \beta < 1  \text{ and } \alpha >0,\\
    \beta = 1 \text{ and } \alpha >2.\\
    \end{cases}
\end{equation}

We now proceed with an estimate of the scaling of $C_L(\infty)$ with $L$.
Assuming the leading trend $b_n \sim n$ in the region $n<n^{*}$ that was conjectured in Ref.~\cite{Parker-19} for generic systems, the first part of the series that appears at the denominator of Eq.~\eqref{Eq:InfiniteTimeFormula} behaves as (see Sec.~\ref{sub:technical} for details)
\be\label{eq:series_1}
\sum^{n^{*}-2}_{n \text{ odd}} \prod^{n}_{n' \text{ odd}}e^{-\Gamma_{n'}} \sim \log n^{*} \sim \log L.
\ee
In order to estimate the remaining part of the series, we first identify the scaling (proven in the \ref{sub:technical})
\be\label{eq:series_2}
\l\frac{b_1b_3\dots b_{n^{*}-2}}{b_2b_4\dots b_{n^{*}-1}}\r^2 \sim \frac{1}{n^{*}} \sim \frac{1}{L}.
\ee
If we assume that for $n>n^{*}$ the rate $\Gamma_n(L)$ is approximately constant and positive  (although small for large $L$), by putting everything together we estimate
\be\label{eq:LGamma}
C_{L}(\infty) \sim \frac{1}{\log L + L^{-1}\sum^{\infty}_{n=0} e^{-n\Gamma(L)}} \sim L\Gamma(L).
\ee
It is now easy to understand the origin of \textit{Conjecture 1}: 
in order to enforce the condition $C_{L}(\infty)\sim L^{-md}$, we need to require $\Gamma(L)\sim L^{-(md+1)}$ as stated in Eq.~\eqref{eq:conj_hydro}.

The \textit{Conjecture 2} concerns those cases where the denominator in Eq.~\eqref{Eq:InfiniteTimeFormula} diverges. 
The formula in Eq.~\eqref{eq:Conj_van}
is proposed to guarantee either a fast approach to $0$ of the rate $\Gamma_n$ or a convergence to a negative value.
By comparing it with the list of the converging cases given in Eq.~\eqref{Eq:Table:Conv}, we see that it encompasses all non-converging situations.

We now consider \textit{Conjecture 3}. 
As we have already discussed, in the presence of strong zero modes, the $b_n$ oscillate between two values and drop to zero after $n \sim L$.
For small $n$, it is thus true that $\Gamma_n$ is constant. The infinite series at the denominator of Eq.~\eqref{Eq:InfiniteTimeFormula} should become a finite sum, whose convergence is automatically assured.
This scenario is compatible with the conjectured inequality~\eqref{eq:conj_zero_mode} because we are essentially saying that $\Gamma_n = + \infty$ for $n$ larger than the termination value.

Let us now discuss the finite-size behavior of the $b_n$ identified for approximate zero modes. 
It is observed numerically that $b_n \sim d_n + (-1)^n c_n$, with $d_n, c_n>0$ and $d_n > c_n$;
it is also shown that $d_n$ is a slowly varying function of $n$ of order $L$ and that $c_n$ drops slowly as a function of $n$~\cite{tmr-25}.
The latter point is crucial:
if the staggering was persistent at large $n$, namely $c_n=c$, 
the rate $\Gamma_n \sim 4c/d_n$ would not decay in $n$ and a non-zero plateau for the autocorrelator would be present, see Eq.~\eqref{Eq:Table:Conv}. 
This scaling is compatible with \textit{Conjecture 3} in Eq.~\eqref{eq:conj_zero_mode};
however, since we are dealing with an approximate zero mode,
we expect that at very large $n$, condition~\eqref{eq:conj_zero_mode} will be violated and the autocorrelation function will decay to zero, after a long and stable plateau.

\subsection{Technical details}\label{sub:technical}

We provide here the technical details about the cumulative products in Eq.~\eqref{eq:cum_prod}.
We first discuss the criterion for the convergence of the cumulative products of Eq.~\eqref{eq:Gamma_family}, and, specifically, we study
\be\label{eq:sum_analysis}
-\log \l\prod^{n}_{n' \text{ odd}} e^{-\Gamma_{n'}} \r = \sum^{n}_{n' \text{ odd}} \Gamma_{n'} \sim \frac{1}{2}\sum^{n}_{n'=1} \frac{\alpha}{(n')^\beta}.
\ee
Here, the factor $1/2$ comes from summing over both even and odd integers, and we only focused on the leading terms for large $n$. Using well-established properties of infinite series, we infer that the sum~\eqref{eq:sum_analysis} converges in the limit $n\rightarrow \infty$ only when $\beta>1$. For $\beta=1$, corresponding to the harmonic series, the leading divergence at large $n$ is
\be
\frac{1}{2}\sum^{n}_{n'=1} \frac{\alpha}{n'} \sim \frac{\alpha}{2}\log n.
\ee
Finally, for $0\leq \beta<1$ a crude estimation gives
\be
\frac{1}{2}\sum^{n}_{n'=1} \frac{\alpha}{(n')^\beta} \sim \frac{\alpha}{2}\int^{n}_1dx \frac{1}{x^\beta} \propto \alpha n^{1-\beta},
\ee
up to a multiplicative constant that is irrelevant for our discussion. Putting everything together and exponentiating the predictions above, we obtain Eq.~\eqref{eq:cum_prod}.

We now discuss the derivation of Eqs.~\eqref{eq:series_1} and \eqref{eq:series_2} related to the values of $b_n$ for $n<n^*$. Assuming $b_n = n$, one obtains $\Gamma_n = -\log (n/(n+1))^2 \simeq 2/n$: this case corresponds to $\beta=1$ and $\alpha=2$ in Eq.~\eqref{eq:Gamma_family}, thus, after exponentiating~\eqref{eq:sum_analysis}, one gets the decay of the cumulative product in Eq.~\eqref{eq:series_2} as $\sim 1/n$. Therefore, the corresponding partial sums grow as $\sim \log n$, as expressed by Eq.~\eqref{eq:series_1}. 

The aforementioned results, which lead to Eq.~\eqref{eq:LGamma}, are derived under the hypothesis that the series of $b_n$ grows perfectly linear in $n$ when $n<n^*$, so the reader might wonder how possible subleading effects (which are known to be present~\cite{Parker-19}) can influence the main conclusions. For instance, recently it has been suggested that the asymptotic scaling of the Lanczos coefficient can be (see Ref.~\cite{lkmv-25})
\be\label{eq:bn_sublead}
b_n \simeq \frac{n}{\log n}\l 1+ \frac{\rho}{2n}\l 1-(-1)^n/\log n\r\r;
\ee
in particular, this is expected for one-dimensional systems in the thermodynamic limit and operators whose spectral density~\eqref{eq:spec_fun} shows the singularity $\Phi(\omega) \sim |\omega|^\rho$ at small $\omega$ (associated with an algebraic decay of the autcorrelator $C(t)\sim t^{-1-\rho}$). A simple calculation shows that~\eqref{eq:bn_sublead} gives the following asymptotics
\be
\l\frac{b_n}{b_{n+1}}\r^2 \simeq 1- \frac{2}{n}+ \frac{2(1+\rho)}{n\log n},
\ee
for odd $n$, and, as a consequence, $\Gamma_n \simeq 2/n$ up to subleading corrections; this is the same behaviour found in the main text to derive Eq.~\eqref{eq:LGamma}, and therefore the corrections in Eq.~\eqref{eq:bn_sublead} are not relevant for our predictions. We understand that as follows: while staggered corrections are present in Eq.~\eqref{eq:bn_sublead}, they decay sufficiently fast in $n$ and do not give rise to non-vanishing plateau in the autocorrelator (in contrast with the scenario discussed in the main text): specifically, although $\Phi(\omega)$ is singular, it does not have $\delta$-singularities, and the corresponding spectrum of $\mathcal{L}$ is continuous.

\section{Numerics}\label{sec:Numerics}

We now present supporting numerical results, obtained with an exact-diagonalization algorithm using up to 600~GB of RAM;
since the computation of the $b_n$ coefficients with the LA is numerically unstable, we resort to the full Gram-Schmidt orthogonalization scheme, which is computationally more expensive.
It is a remarkable fact discussed in Appendix~\ref{app:numerics} that the coefficients computed with the standard LA algorithm describe the same autocorrelation function and satisfy the same universal scaling laws that we present here below.

\subsection{Simple checks: short-range one-dimensional models}
In this section, we focus on the model in Eq.~\eqref{Eq:TFIM}; additional numerical studies addressing long-range systems and higher-dimensional lattices are presented in Sec.~\ref{sub:numerics}, supporting our conjecture and its claimed universality.
We consider the three observables $\mathcal{O} = \sigma^z_1$, $\sigma^z_1\sigma^z_2$ and $\sigma^y_1$. 
According to the analysis presented in Ref.~\cite{cwxmp-25}, $\sigma^z_1$ overlaps (in the infinite temperature state) with $H$ and $\sigma^z_1\sigma^z_2$ with $H^2$: as a consequence, the plateaus $C_{L}(\infty)$ scale respectively as $L^{-1}$ and $L^{-2}$. Hence, according to the \textit{Conjecture~1}, $\Gamma_n(L)$ is expected to approach a positive constant $\sim L^{-(m+1)}$, with $m=1,2$ respectively. 

\begin{figure}[t]
\includegraphics[width=\columnwidth]{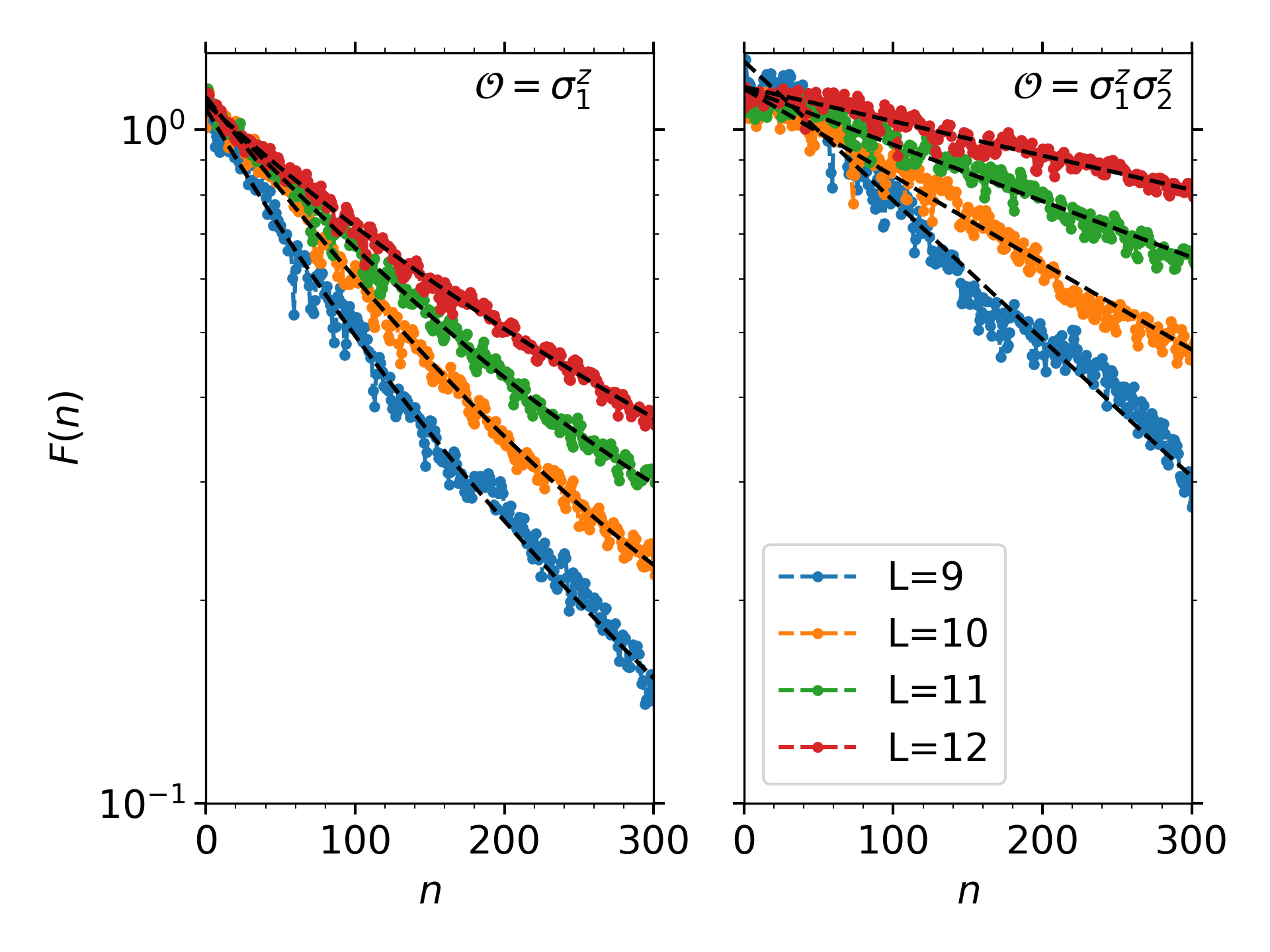}
\caption{The cumulative product $F(n)$ for the operators $\sigma^z_1$ (left panel) and $\sigma^z_1\sigma^z_2$ (right panel). The data refer to the Ising Hamiltonian \eqref{Eq:TFIM} with parameters $[J,h_x,h_z] = [1,1,1.5]$, and the dashed black lines show the fit with bi-exponential curves.}
\label{Fig:TFIM:Ratios:conj1}
\includegraphics[width=\columnwidth]{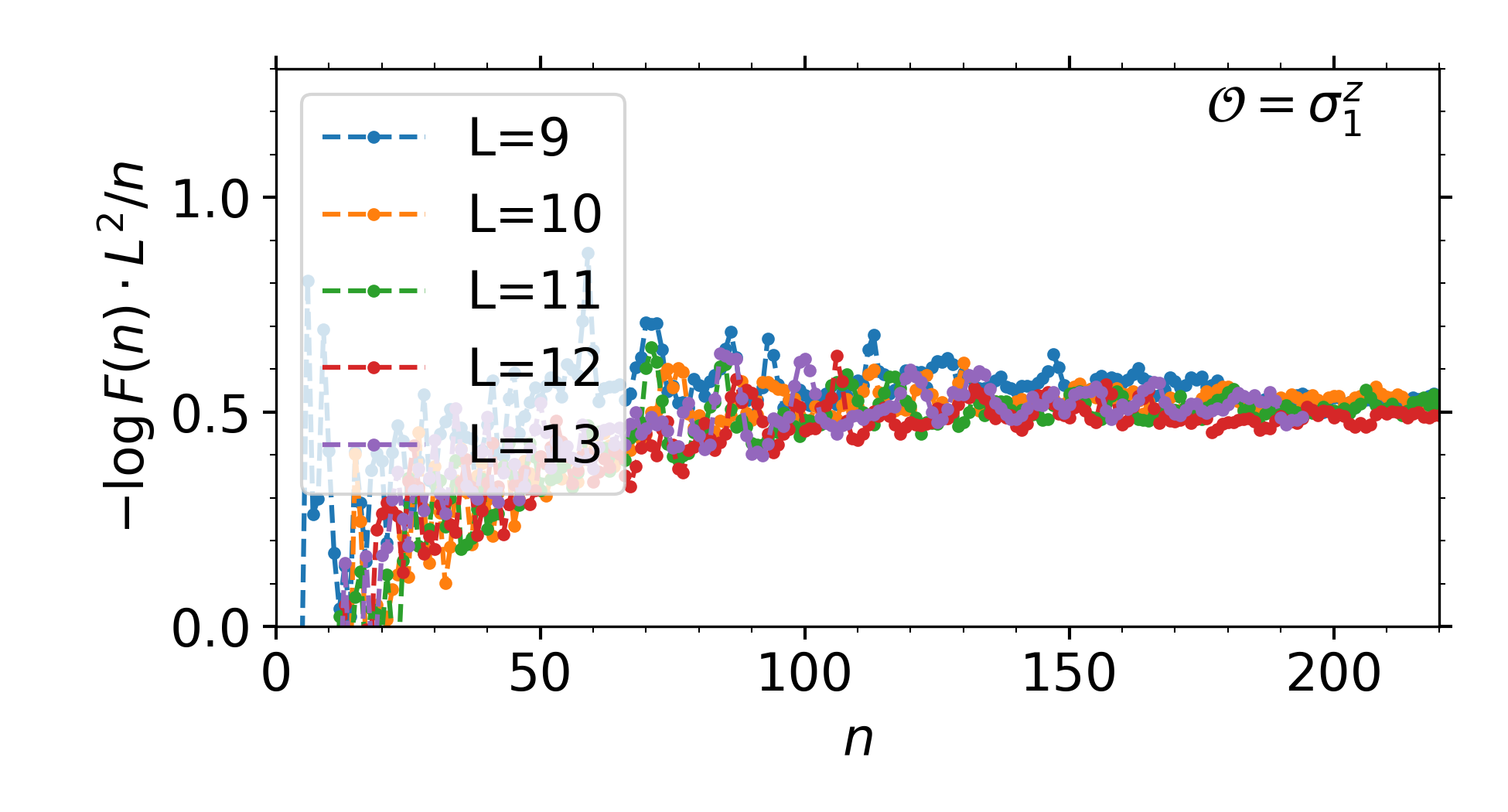}
\caption{The cumulative products in Fig.~\ref{Fig:TFIM:Ratios:conj1} for the operator $\sigma^z_1$ ($m=1$) rescaled as $-\log F(n)\cdot L^{m+1}/n$. 
Assuming conjecture \eqref{eq:conj_hydro}, the curves are expected to collapse for large $L$.}
\label{Fig:TFIM:Ratios:conj1_bis}
\end{figure}

To check whether this prediction has some descriptive power, we plot in Fig.~\ref{Fig:TFIM:Ratios:conj1} the cumulative products
\be\label{eq:cum_prod1}
F(n) := \prod^{n}_{n'= 0}e^{-\Gamma_{n^*+2n'}(L)}.
\ee
Here, $n^{*}$ is chosen as the first odd number that is found after $b_n$ reaches its maximum in the crossover regime, see Fig.~\ref{Fig:1:Example}. 
If $\Gamma_n$ were perfectly constant, $F(n)$ would decay exponentially in $n$. Although persistent fluctuations and possible subleading corrections are present around the leading trend, the decay is visible. We fit $F(n)$ as a bi-exponential, assuming $F(n)\simeq a_1 e^{- c_1 n}+ a_2 e^{- c_2 n}$, identifying an averaged rate $\bar{\Gamma} := (a_1/c_1+a_2/c_2)^{-1}$.

For both operators, the data are consistent with the presence of a decay rate that decreases with the system size. 
In order to check whether the scaling of the rate with $L$ is compatible with Eq.~\eqref{eq:conj_hydro}, in Fig.~\ref{Fig:TFIM:Ratios:conj1_bis} we plot the quantity $-\log F(n)\cdot L^{m+1}/n$ as a function of $n$ for different sizes: these curves are expected to collapse for large $L$ toward a constant value. 
The results for $\sigma^z_1$ in Fig.~\ref{Fig:TFIM:Ratios:conj1_bis} display a good agreement with the conjecture; the data for $\sigma^z_1\sigma^z_2$ are less convincing, and a better agreement would be found if larger values of $n$ were accessible: we comment on the issue in the Appendix~\ref{app:numerics}. 

\begin{figure}[t]
\includegraphics[width=\columnwidth]{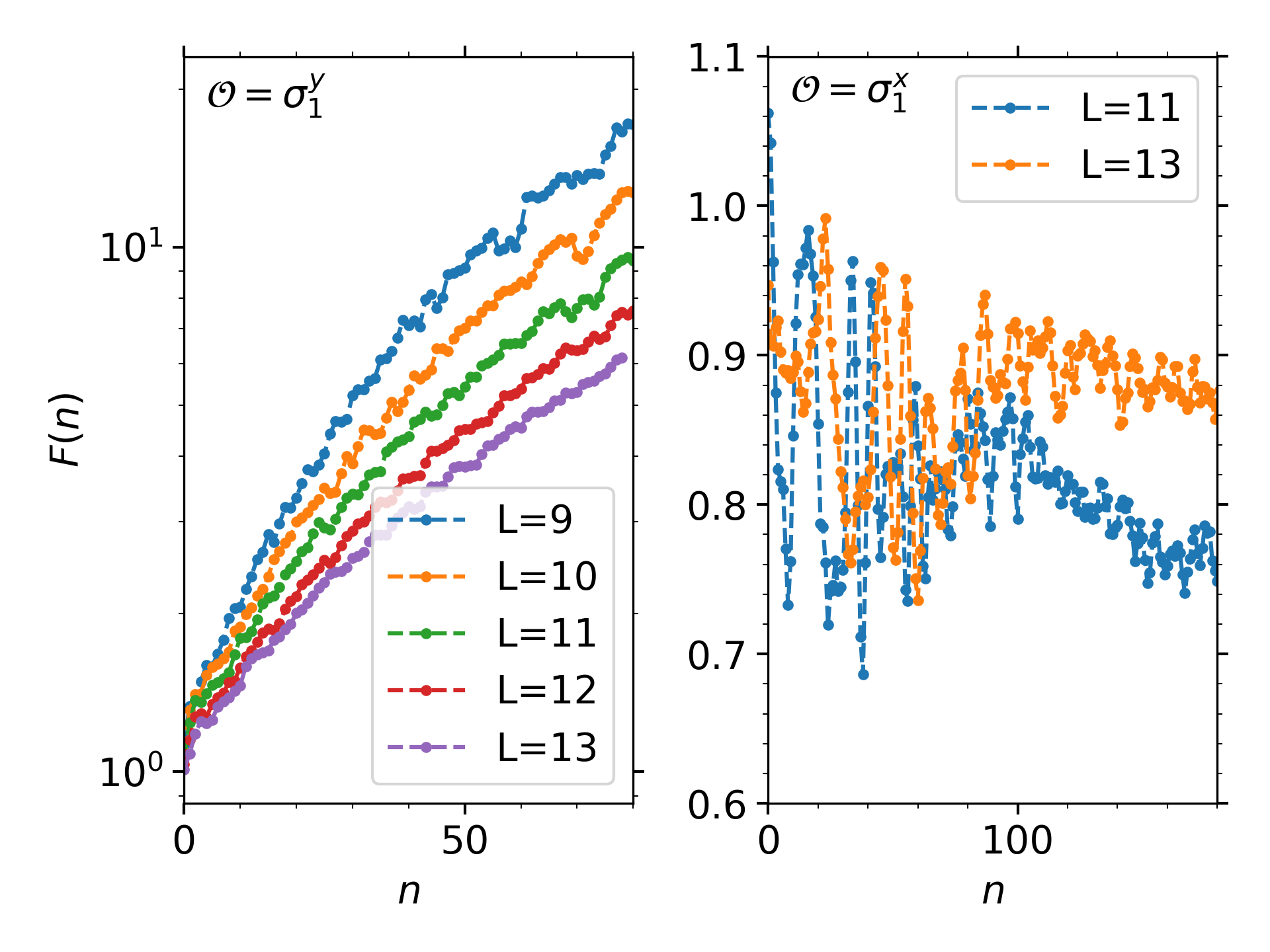}
\caption{(Left) The cumulative product $F(n)$ for the Ising model~\eqref{Eq:TFIM} and the operator $\sigma^y_1$: the data shows a monotonic increasing trend. 
(Right) The cumulative product $F(n)$ for Hamiltonian~\eqref{eq:H_zm}, hosting an approximate zero mode, and the operator $\sigma^x_1$: the data are consistent with saturation to a finite value.}
\label{Fig:TFIM_conj_van}
\end{figure}

We now move to the study of the operator $\sigma^y_1$, which does not overlap with any power of $H$ (being odd under time reversal)~\cite{cwxmp-25}; 
the infinite-time plateau value of $C_{L}(\infty)$ is expected to vanish and thus it allows to test \textit{Conjecture 2}. Our numerics is presented in Fig.~\ref{Fig:TFIM_conj_van} (left panel), where we show that in this case $F(n)$ is an increasing function of $n$. This means that $\Gamma_n(L)$ fluctuates around a negative value, and thus is compatible with the inequality~\eqref{eq:Conj_van}, which summarizes \textit{Conjecture 2}.

Finally, we consider the Hamiltonian
\be\label{eq:H_zm}
H = -\sum_{i=1}^{L-1}\sigma^x_i\sigma^x_{i+1}+U\sum_{i=1}^{L-1}\sigma^z_i\sigma^z_{i+1}-\frac{\mu}{2}\sum_{i=1}^L\sigma^z_i
\ee
with open boundary conditions and parameters $\mu=0.2$, $U=0.5$; the model is known to host a boundary zero mode at zero temperature that becomes approximate when the temperature is increased~\cite{tmr-25}. We probe the operator $\mathcal{O} = \sigma^x_1$ at the left boundary of the chain in the infinite temperature state. 
We show the results in Fig.~\ref{Fig:TFIM_conj_van} (right panel): the data for $F(n)$ are compatible with a saturation at large $n$, consistent with $\Gamma_n \simeq 0$, implying $C_L(\infty)=0$ according to~Eq.~\eqref{Eq:Table:Conv}. 
If the zero mode were exact, we would have observed an exponential decay of $F(n)$ with a size-independent decay rate $\Gamma$. Our data are thus compatible with \textit{Conjecture 2}, although from these numerical data it is difficult to extract reliable quantitative values for the large $n$ scalings of $\Gamma_n(L)$. By comparing the two panels of Fig.~\ref{Fig:TFIM_conj_van},
we observe a marked qualitative difference between the two cases, which share the common property $C_L(\infty)=0$, but are associated to two qualitatively distinct physical situations (absence of hydrodynamic projection versus approximate zero mode).

\subsection{Further numerical benchmarks: long-range and higher dimensional systems}\label{sub:numerics}

\begin{figure}[t]
\includegraphics[width=\columnwidth]{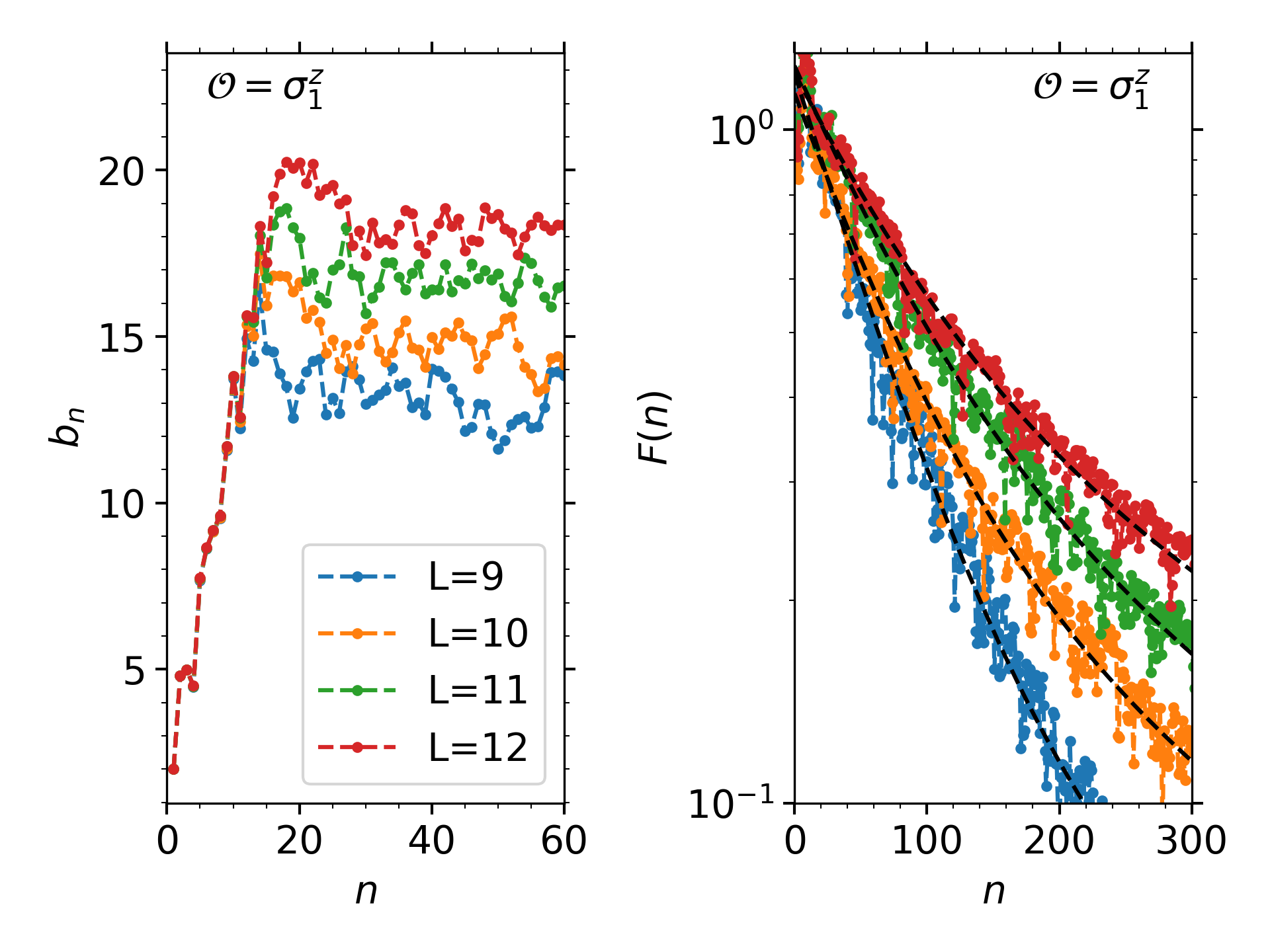}
\caption{(Left) Lanczos coefficients $b_n$ associated with the operator $\sigma^z_1$ for the long-range Ising Hamiltonian in Eq. \eqref{eq:Hlongrange} with $\alpha=1.5, h_z=1.5$. (Right) The function $F(n)$ for different system sizes; the colored dots are numerical data, while the dashed black lines are biexponential fits.}
\label{Fig:LR_Z}
\end{figure}

We consider a one-dimensional Ising model with long-range interactions:
\be\label{eq:Hlongrange}
H = \sum_{i=1}^L\sum_{j>i}\frac{1}{N_{\alpha}r_{ij}^{\alpha}}\sigma^x_i\sigma^x_j+\sum_{i=1}^L h_z\sigma^z_i
\ee
where 
\be 
r_{ij}=\mathrm{min}(|j-i|,L-|j-i|), \quad N_{\alpha}=\left(\sum_{i=2}^L\frac{1}{r_{i1}^{2\alpha}}\right)^{1/2},
\ee
and we focus on the operator $\mathcal{O}=\sigma^z_1$. 
The computed values of $b_n$ are shown in Fig.~\ref{Fig:LR_Z} (left).
Here, since this model is long-ranged, the values of $b_n$ do not coincide exactly across different sizes for sufficiently small $n$, unlike in short-range systems (see, for instance, Fig.~\ref{Fig:1:Example}). Nonetheless, a qualitative agreement is observed, and for $n>n^*$ the Lanczos coefficients saturate to a plateau. Qualitatively, thus, we have the same behavior displayed in Fig.~\ref{Fig:1:Example}.

We then analyse the $b_n$ coefficients and in Fig.~\ref{Fig:LR_Z} (right) we plot $F(n)$ in Eq.~\eqref{eq:cum_prod1}, choosing $n^*$ with the same convention as in the short-range case, and fit the data with a bi-exponential, obtaining an excellent agreement. 
As in the short-range model, the long-range system shows an asymptotic exponential decay of $F(n)$, with a decay rate that diminishes as $L$ increases.

However, in the presence of long-range interactions, we have no reason to expect $n^{*}$ to scale as $\sim L$, and therefore we cannot directly employ the prediction of Eq.~\eqref{eq:conj_hydro}. To the best of our knowledge, this problem has never been studied theoretically in the literature, and the accessible sizes are not sufficiently large to propose a convincing ansatz for the behavior of $n^{*}$ in long-range models based on numerical data. Nonetheless, following the steps leading to Eq.~\eqref{eq:LGamma} and assuming $b_n$ is linear in the initial regime, we deduce
\be
C_L(\infty) \sim n^{*}\Gamma(L).
\ee
Thus, if $C_L(\infty) \sim 1/L^m$ (here we expect $m=1$, even if the model is long-range; see Ref. \cite{cwxmp-25} for details) and $n^{*}\sim L^a$ for some positive value of $a$, we can still infer an algebraic decay of $\Gamma(L)$ as
\be
\Gamma(L)\sim L^{-m-a}.
\ee

\begin{figure}[t]
\includegraphics[width=\columnwidth]{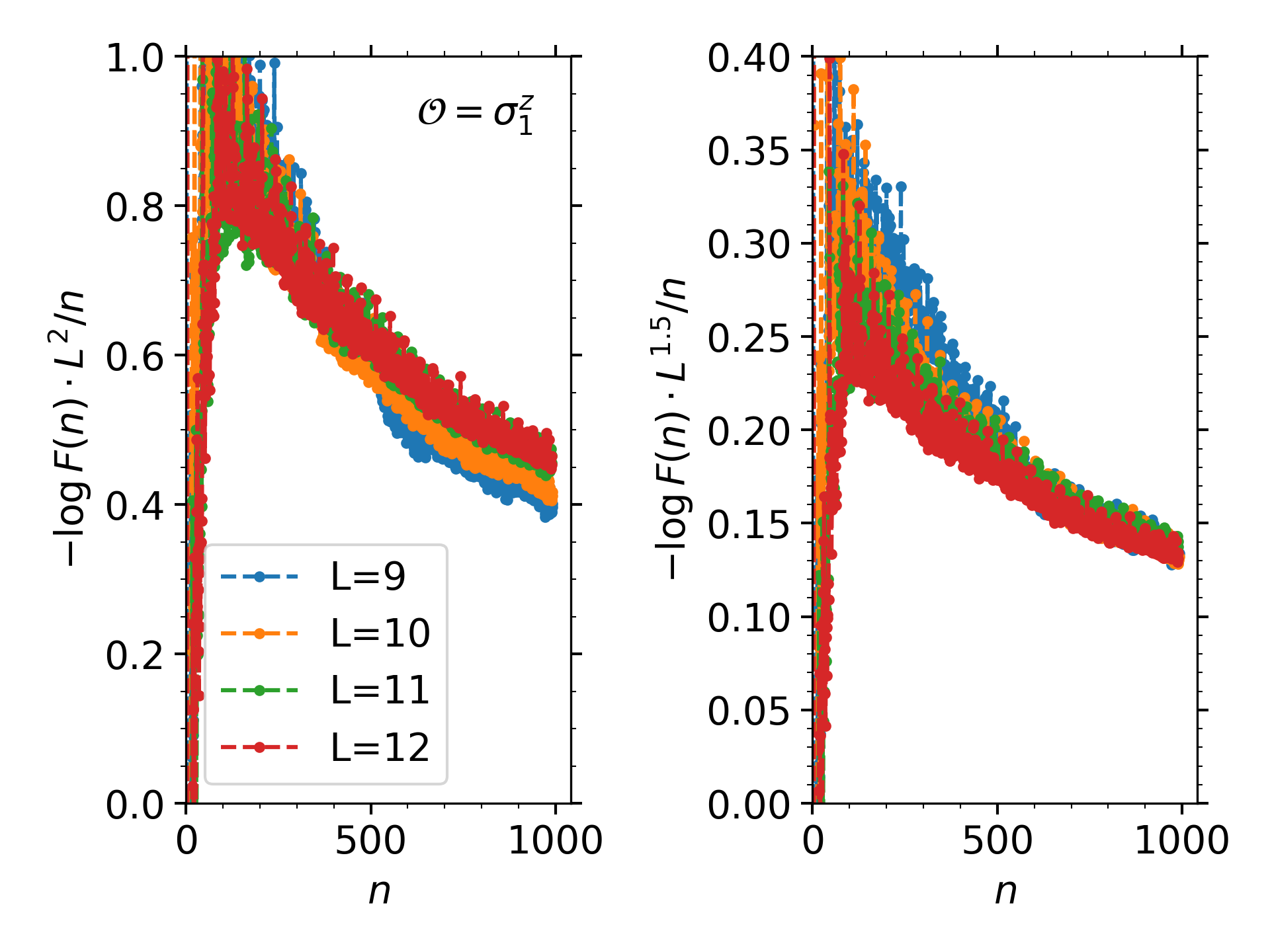}
\caption{We plot the quantity $-L^{1+a}\log F(n) /n$ for the operator $\sigma^z_1$: we consider $a=1$ (left) and $a=0.5$ (right). The data refer to the long-range Ising Hamiltonian in Eq. \eqref{eq:Hlongrange} with $\alpha=1.5, h_z=1.5$.
}
\label{Fig:LR_Z_Lscaling}
\end{figure}

In Fig.~\ref{Fig:LR_Z_Lscaling}, we test the compatibility of our data with an algebraic decay $\Gamma(L)$ by plotting the function $-\log F(n)L^{1+a}/n$. In the left panel, we consider $a=1$, which corresponds to the scaling expected from Eq. \eqref{eq:conj_hydro}, and namely with \textit{Conjecture 1}. In the right panel, we choose $a=0.5$, a value that exhibits better compatibility with a collapse. Although we cannot make strong claims about the precise value of $a$, it is reasonable to expect $a<1$, since long-range interactions tend to accelerate the spreading of operators, thereby decreasing $n^{*}$ compared to their short-range counterparts.

\begin{figure}[t]
\includegraphics[width=\columnwidth]{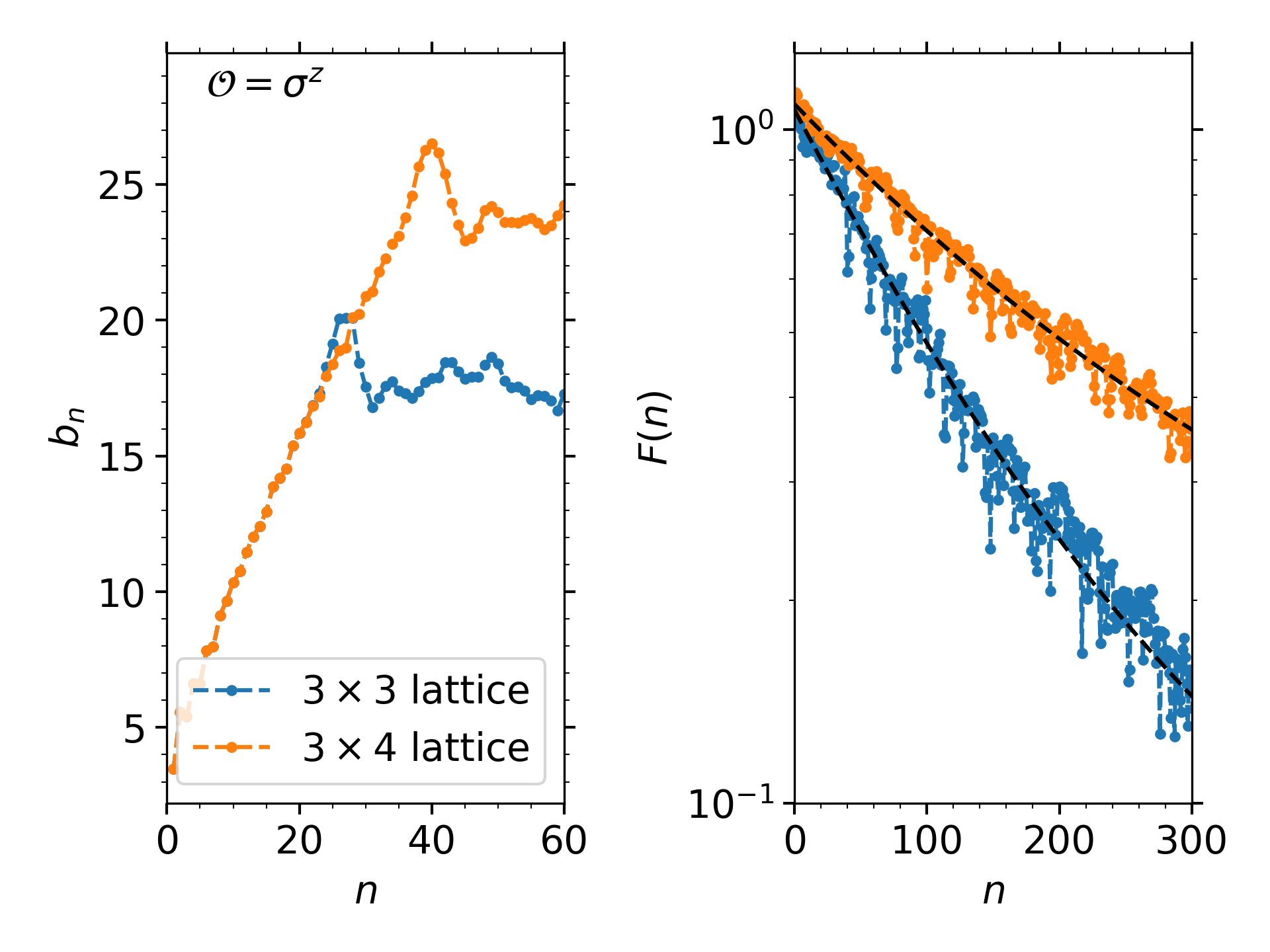}
\caption{
We plot the Lanczos coefficients $b_n$ (left panel) and $F(n)$ (right panel) for the operator $\mathcal{O}=\sigma^z_{(2,2)}$. The data refer to 2D Ising Hamiltonian in Eq. \eqref{eq:H2d} with parameters $[J,h_x,h_z] = [1,1,1.5]$ and lattice sizes $(L_x,L_y) = (3,3),(3,4)$.}
\label{Fig:2d_Z}
\end{figure}

We also consider a two-dimensional short-range model
\be\label{eq:H2d}
H =  \sum_{\langle \mathbf{i},\mathbf{j}\rangle}J  \sigma_\mathbf{i}^x  \sigma_{\mathbf{j}}^x
 + \sum_\mathbf{j}(h_x  \sigma_\mathbf{j}^x + h_z  \sigma_\mathbf{j}^z ),
\ee
where $\langle \mathbf{i},\mathbf{j}\rangle$ denotes the nearest neighbours on a $L_x\times L_y$ lattice with open boundary conditions. We consider $(L_x,L_y) = (3,3),(3,4)$ and we focus on the operator $\sigma^z_{(2,2)}$ (inserted at the center of the $3\times 3$ lattice). In Fig. \ref{Fig:2d_Z} (left), we show the Lanczos coefficients $b_n$: they display an initial linear growth, $b_n\sim n$, and a plateau for $n>n^{*}$, resembling the behaviour observed for one-dimensional systems. Similarly, we show the decay of the function $F(n)$ in Fig. \ref{Fig:2d_Z} (right), and the numerical data are compatible with a bi-exponential decay. 

In this case, the computational cost limits access to larger system sizes with standard numerics, making the finite-size scaling analysis difficult. Nonetheless, since the value of the plateau is expected to scale as $1/L^d$ ($d=2$, and $L$ the typical size of the lattice) \cite{cwxmp-25}, and the expected value of $n^*$ is $n^*\sim L$, as a consequence of locality of the interactions, we predict $\Gamma(L)\sim L^{-3}$ compatible with the \textit{Conjecture 1} and specifically with Eq.~\eqref{eq:conj_hydro}. This is consistent with numerical data, since we observe that the rate $\Gamma_n(L)$ is smaller when the lattice is bigger. 

\section{Conclusions and outlook}\label{sec:Conclusions}
We considered the LA for quantum many-body systems, and we focused on the Lanczos coefficients $b_n$, computing several hundred of them for finite-size systems. 
The universal properties of autocorrelation functions in generic models allowed us to present a threefold conjecture on the scaling of the ratios $(b_n / b_{n+1})^2$, with $n$ odd, in the large $n$ limit.
Our main goal was to show that the finite-size behavior of the Lanczos coefficients has universal properties that can be studied with state-of-the-art numerics.

This work opens intriguing research directions. 
A natural question is to understand the origin of the bi-exponential behavior found in Fig.~\ref{Fig:TFIM:Ratios:conj1}.
Moreover, after having studied the late-time plateau of $C_L(t)$, it is natural to focus on the algebraic decay $t^{-\nu}$ (partially addressed in Ref.~\cite{lkmv-25}), and to investigate how the finite-size Lanczos coefficients give rise to such a behavior. 
Also, the precise mechanism by which the finite-size crossover of $C_L(t)$, from decay to plateau, occurs at the level of the LA is currently unclear.
Finally, our study of the Lanczos coefficients has highlighted an erratic behavior in the region $n>n^{*}$; whether the latter has some universal properties, and it can be studied as a stochastic process, is a question that deserves to be addressed.

\acknowledgments

We warmly thank X.~Cao for enlightening discussions and for sharing with us his experience on the subject.
We thank B.~Buca, A.~Mitra, M.~Rizzi and N.~Tausendpfund for discussions on the subject.
L.C.~and L.M.~thank J.~Wang, X.~Xu and D.~Poletti for related discussions during a previous collaboration.
This work is supported by
the ANR project LOQUST ANR-23-CE47-0006-02.

\begin{appendix}

\section{On the numerical instability of the Lanczos algorithm and its effect on the universal properties of the Lanczos coefficients at finite size}\label{app:numerics}

The focus of this work is the universal properties of the Lanczos coefficients $b_n$ at finite size $L$, which are produced as follows. The Lanczos algorithm (LA) starts with a seed operator $O_0=\mathcal{O}$, and then builds an orthonormal basis of operators $\{O_n\}$ by repeatedly acting with the Liouvillian $\mathcal{L}$ to $O_0$ while orthonormalizing against the one obtained at the step before. In formulas, this means that, at each iteration, we compute 
\begin{equation}\label{eq:3steps}
\begin{split}
    O'_n&=\mathcal{L}O_{n-1}-b_{n-1}O_{n-2},\\
    O_n&=\frac{O'_n}{b_n},\\
    b_n&=||O'_n||,
\end{split}
\end{equation}
with $b_0=0$ and $||\mathcal{O}||=\sqrt{\la \mathcal{O}^\dagger\mathcal{O}\ra}$. This implies that at any iteration of the algorithm, only three elements of the basis $\{O_n\}$ must be stored in memory and permits the computation of several thousands of $b_n$ for chain lengths up to $L=14$. 
However, as we have emphasized in the main text, as we increase $n$, the LA may fail to produce perfectly orthogonal vectors (a discussion of this known fact can be found, for instance, in Ref.~\cite{Rabinovici2023}). This means that, unless we store more than 3 vectors, we cannot guarantee that our Krylov-basis $\{O_n\}$ is orthonormal. To correct for orthogonalization errors, we keep every computed state in memory and apply a Gram–Schmidt procedure to each newly generated vector.
However, storing all $O_n$ requires a large amount of memory, and $L=13$ is the largest system size for which we can perform the extra Gram-Schmidt orthogonalization step. 

\begin{figure}[h]
  \centering
    \includegraphics[width=0.7\linewidth]{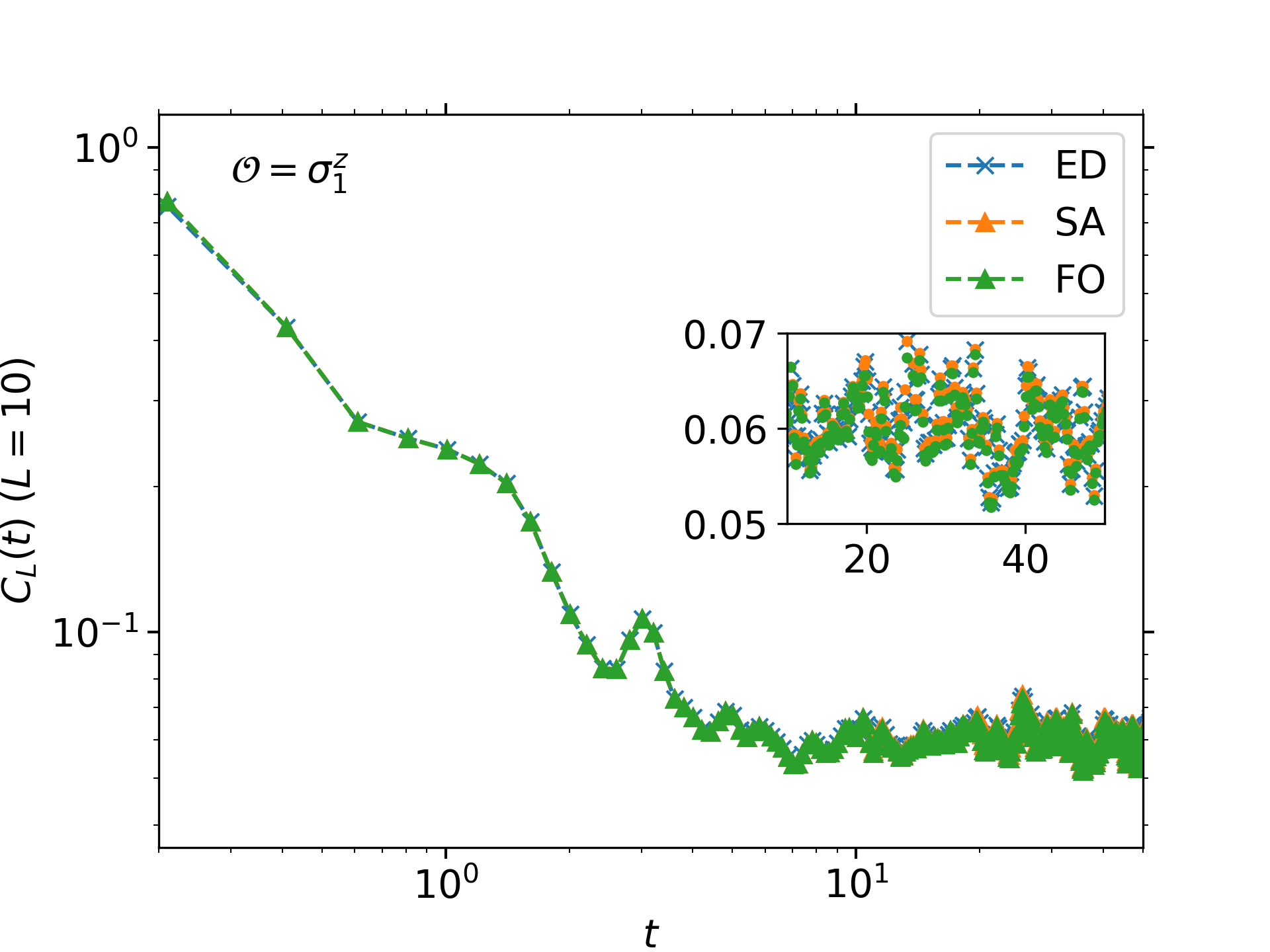}
    \includegraphics[width=0.7\linewidth]{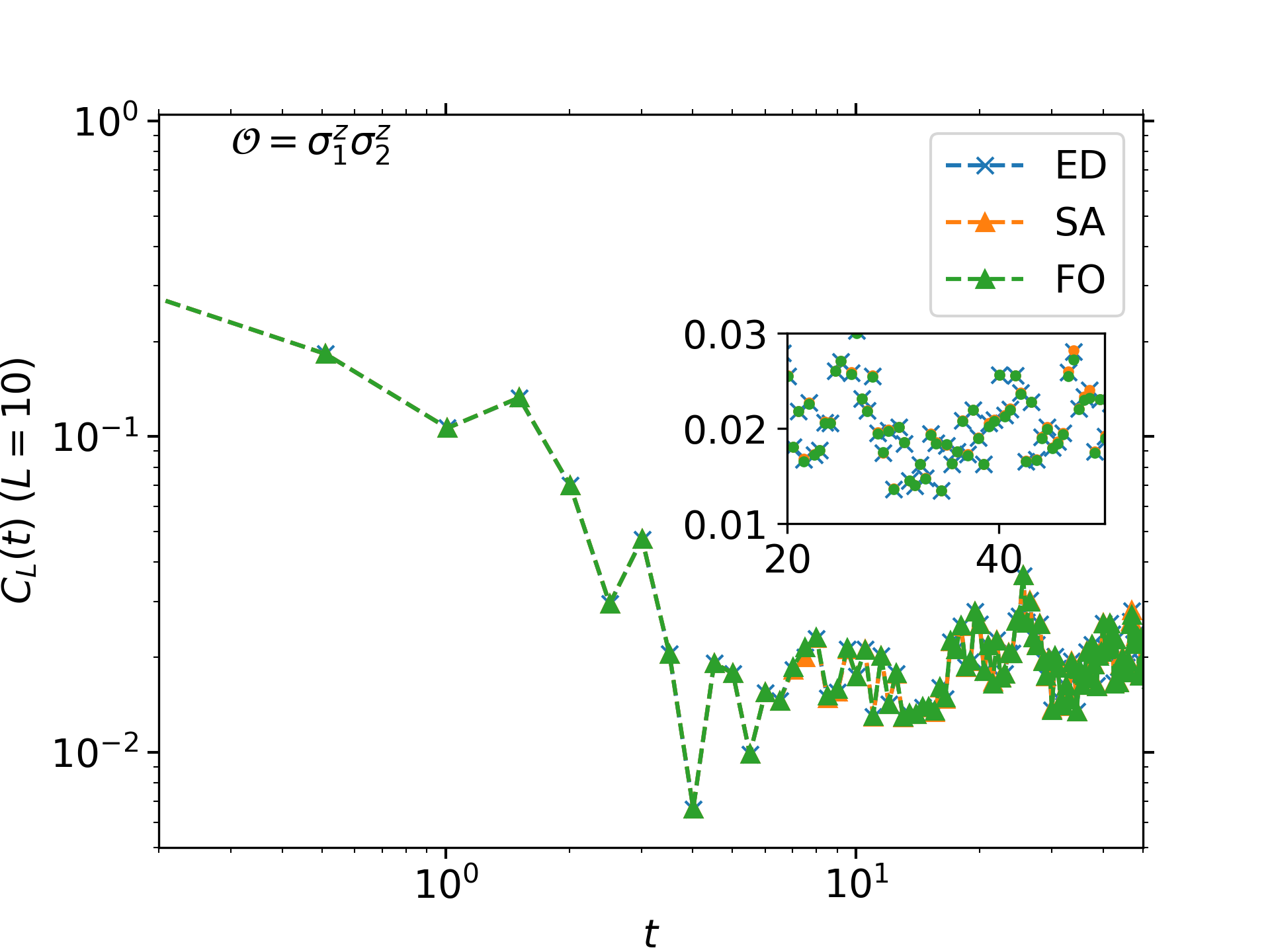}
    
 %
%
%
    \includegraphics[width=0.7\linewidth]{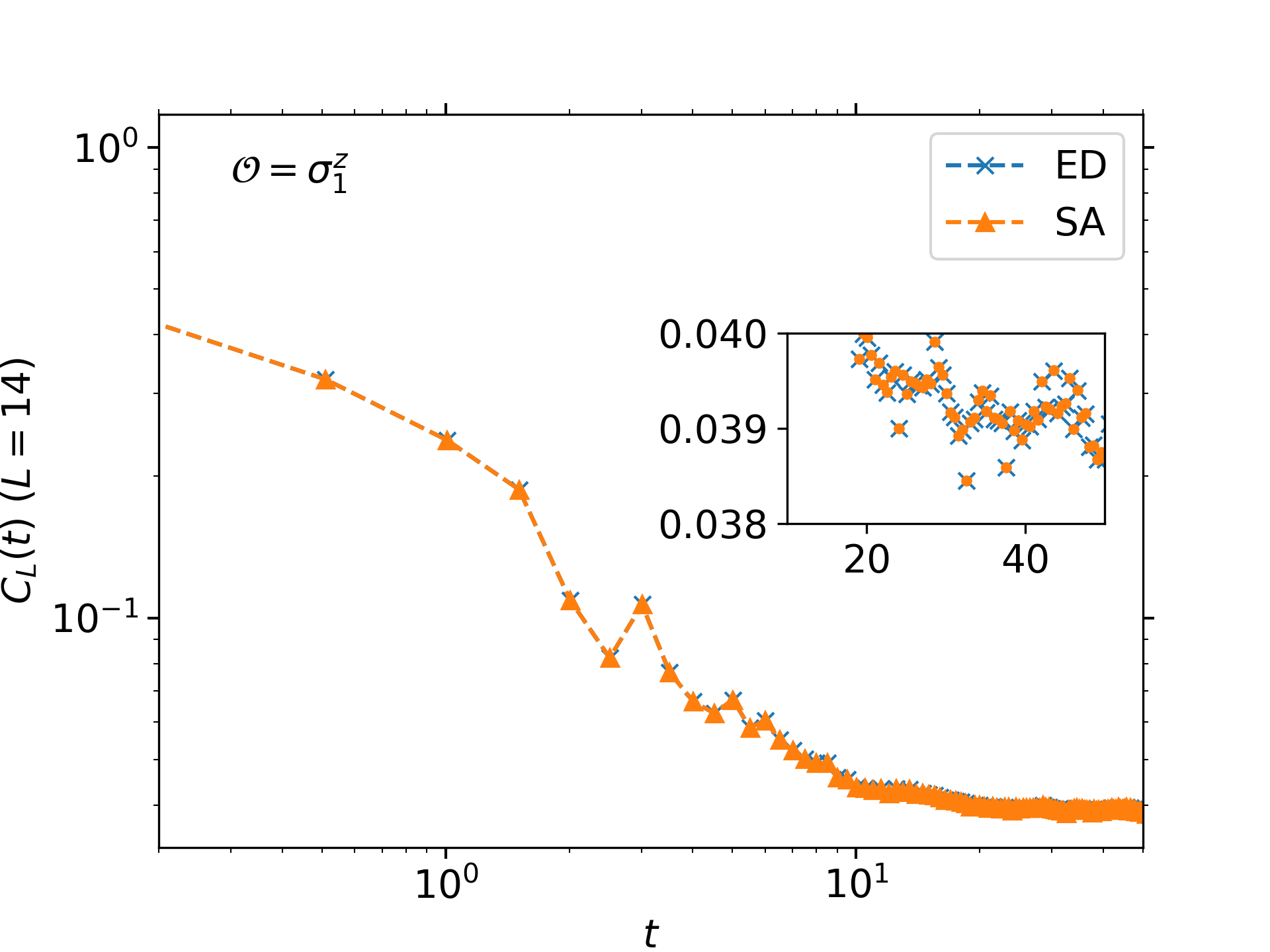}
   

  \caption{Time evolution of the autocorrelation $\langle \mathcal{O}(t)\mathcal{O}\rangle$ function for the Ising model in Eq.~(\textcolor{blue}{3}) of the main text. In the top and central panels, $\mathcal{O}=\sigma_1^z$, while in the bottom $\mathcal{O}=\sigma_1^z\sigma_2^z$. The different colors correspond to different methods: the blue symbols have been obtained using Exact Diagonalization (ED), the orange lines are based on the standard Lanczos algorithm (SA) described in Eqs.~\eqref{eq:3steps}, while the green symbols are obtained using the LA with the full Gram-Schmidt orthogonalization (FO). The LA has been truncated at $n=1000$ for all the autocorrelators at $L=10$, while for $L=14$ the truncation occurs at $n=2000$. The plots clearly show that both algorithms reproduce the same autocorrelations despite the numerical instabilities of the SA.}\label{fig:Aut_Z}

\end{figure}

\begin{figure}[t]
\includegraphics[width=\columnwidth]{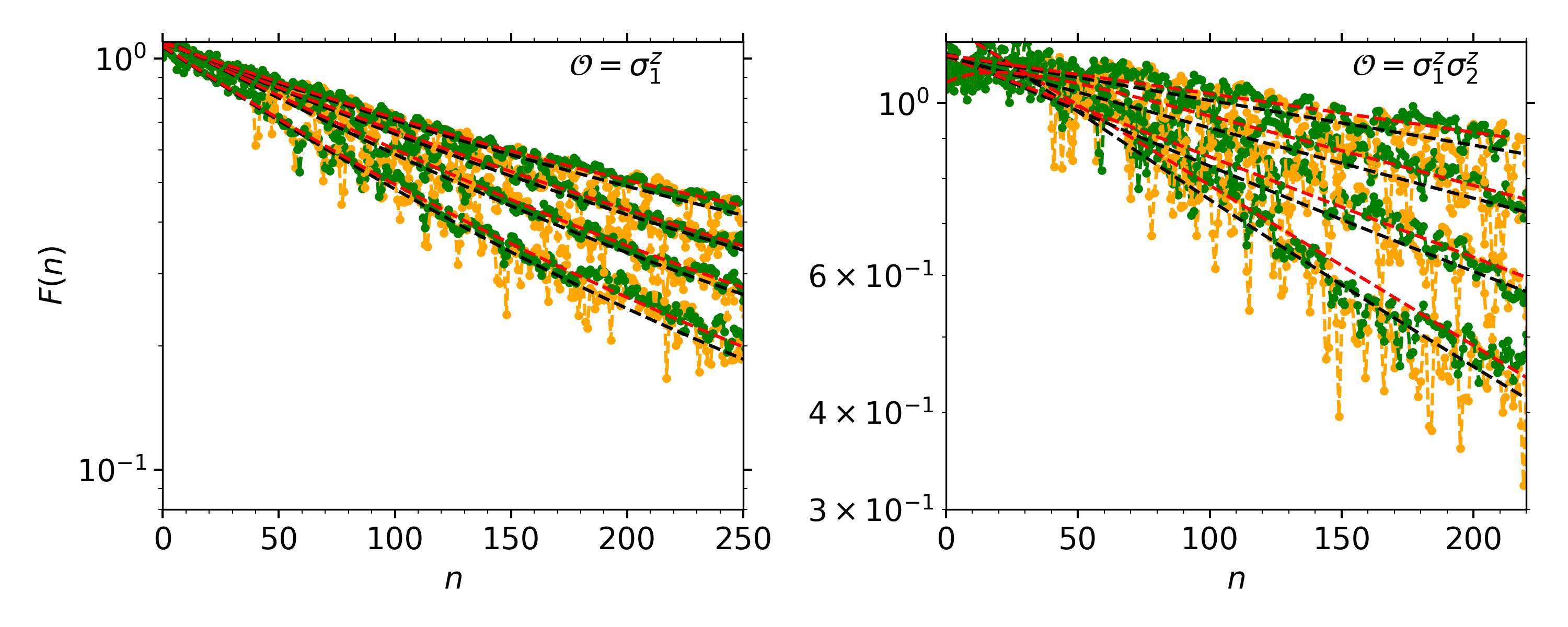}
\caption{Comparison between the cumulative products $F(n)$ obtained through the 3-step algorithm of Eq.~\eqref{eq:3steps} (yellow) and with the full orthogonalization (green) for different system sizes $L=9,10,11,12$ (bottom to top). The dashed lines correspond to bi-exponential fits for the SA data (black) and the FO ones (red). Both panels show a good qualitative and quantitative agreement between the parameters extracted from the fit.}
\label{Fig:Fn_comp}

\end{figure}

Given these two methods, one can ask what is the impact of the numerical instabilities of the Lanczos coefficients in the quantities we are interested in here, namely the autocorrelation function in Eq.~(\textcolor{blue}{1}) of the main text, and the cumulative products in Eq.~(\textcolor{blue}{15}) of the main text. 
To address this question, we compare in Fig.~\ref{fig:Aut_Z} the behaviour of the autocorrelations of $\mathcal{O}=\sigma^z_1$ and $\sigma^{z}_1\sigma_1^z$ calculated by exact diagonalization (ED) with those computed using the LA both with (FO, full orthogonalization) and without (SA, standard algorithm) the full Gram-Schmidt orthogonalization. 
With FO, the maximum value of $n$ we can reach for $L=10$ is $n=1000$. For this reason, we restrict to the same window for the three-step standard algorithm of Eq.~\eqref{eq:3steps} for $L=10$, while we reach $n=2000$ for $L=14$. In the three plots of Fig.~\ref{fig:Aut_Z}, we clearly observe that the discrepancies arising from the imperfect orthogonalization procedure does not have any strong impact on the time-evolution of the autocorrelations, which is the main focus of this work. The deviations from the exact results are $O(10^{-4})$ with both approaches, also at larger system sizes.


\begin{figure}[t]
    \includegraphics[width=0.7\linewidth]{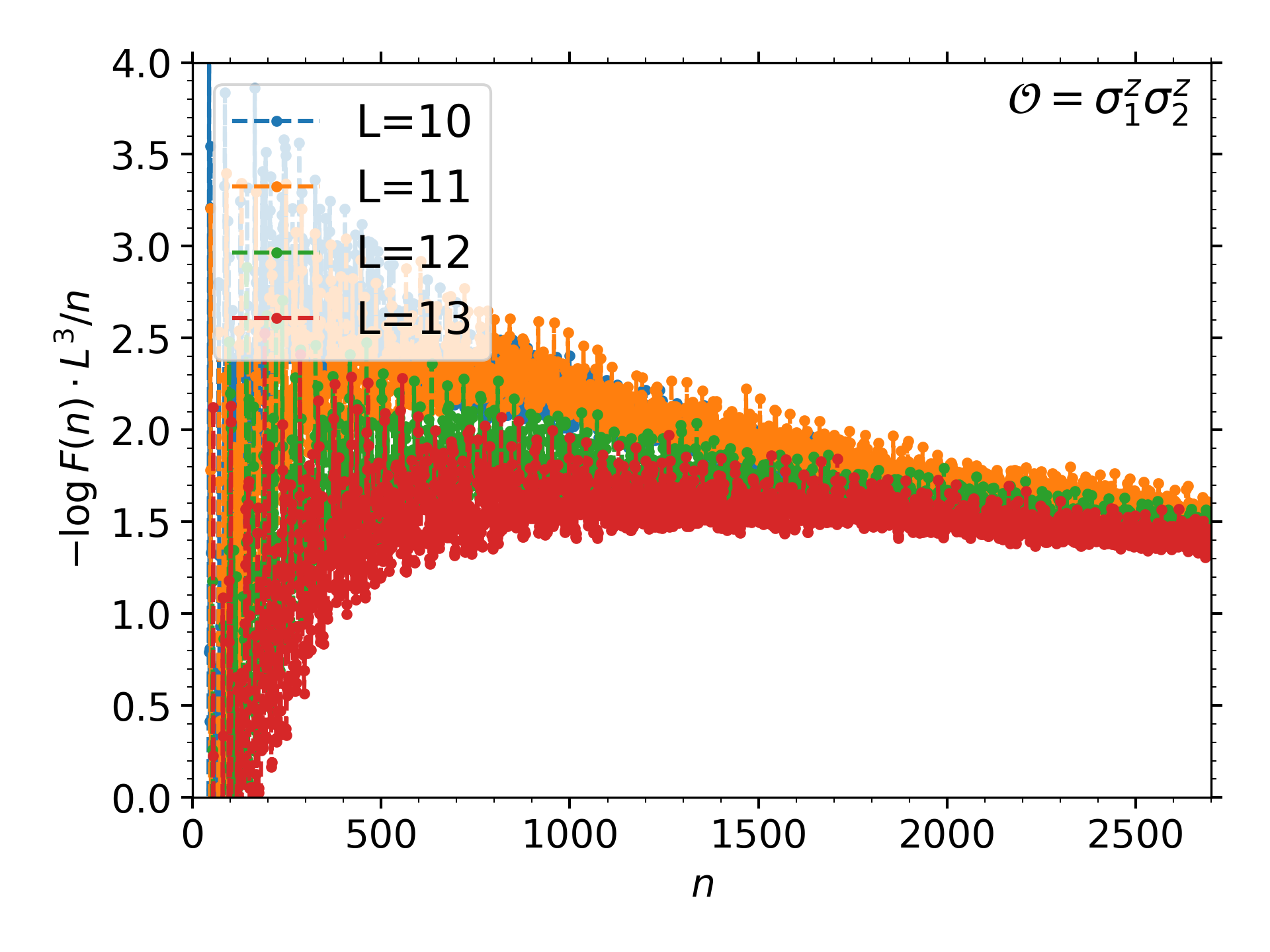}

    \includegraphics[width=0.7\linewidth]{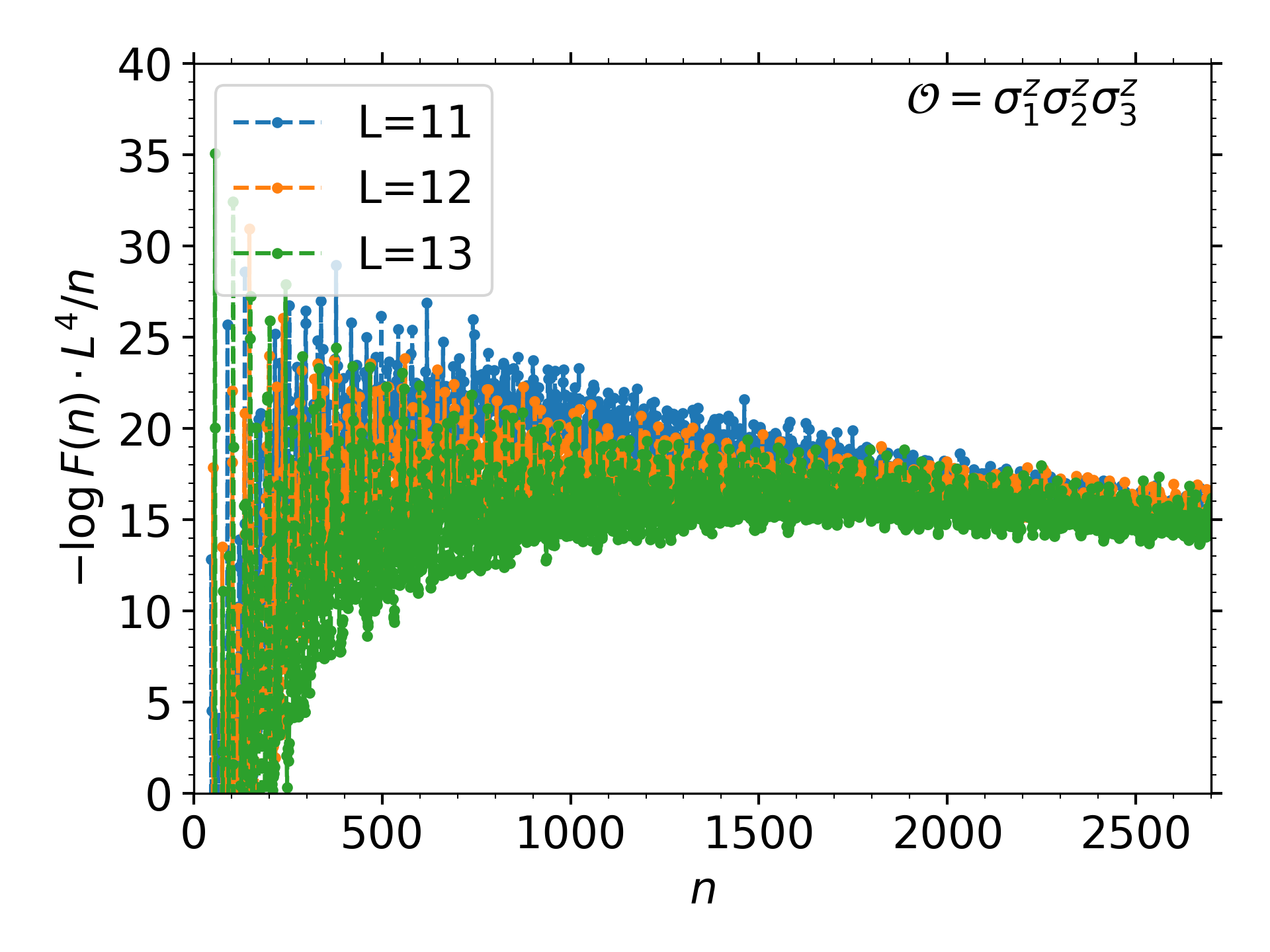}
   
    \caption{Behavior of $-\log F(n)\cdot L^{m+1}/n$ for the operators $\mathcal{O}=\sigma_1^z\sigma_2^z$ ($m=2$, top panel) and $\mathcal{O}=\sigma_1^z\sigma_2^z\sigma_3^z$ ($m=3$, bottom panel). Using $6000$ Lanczos coefficients $b_n$, the different curves collapse at large values of $n$, confirming the validity of our first conjecture, summarized by Eq.~(\textcolor{blue}{6}) in the main text.}
    \label{fig:Fn_long}
\end{figure}

\begin{figure}[t]
\includegraphics[width=0.7\columnwidth]{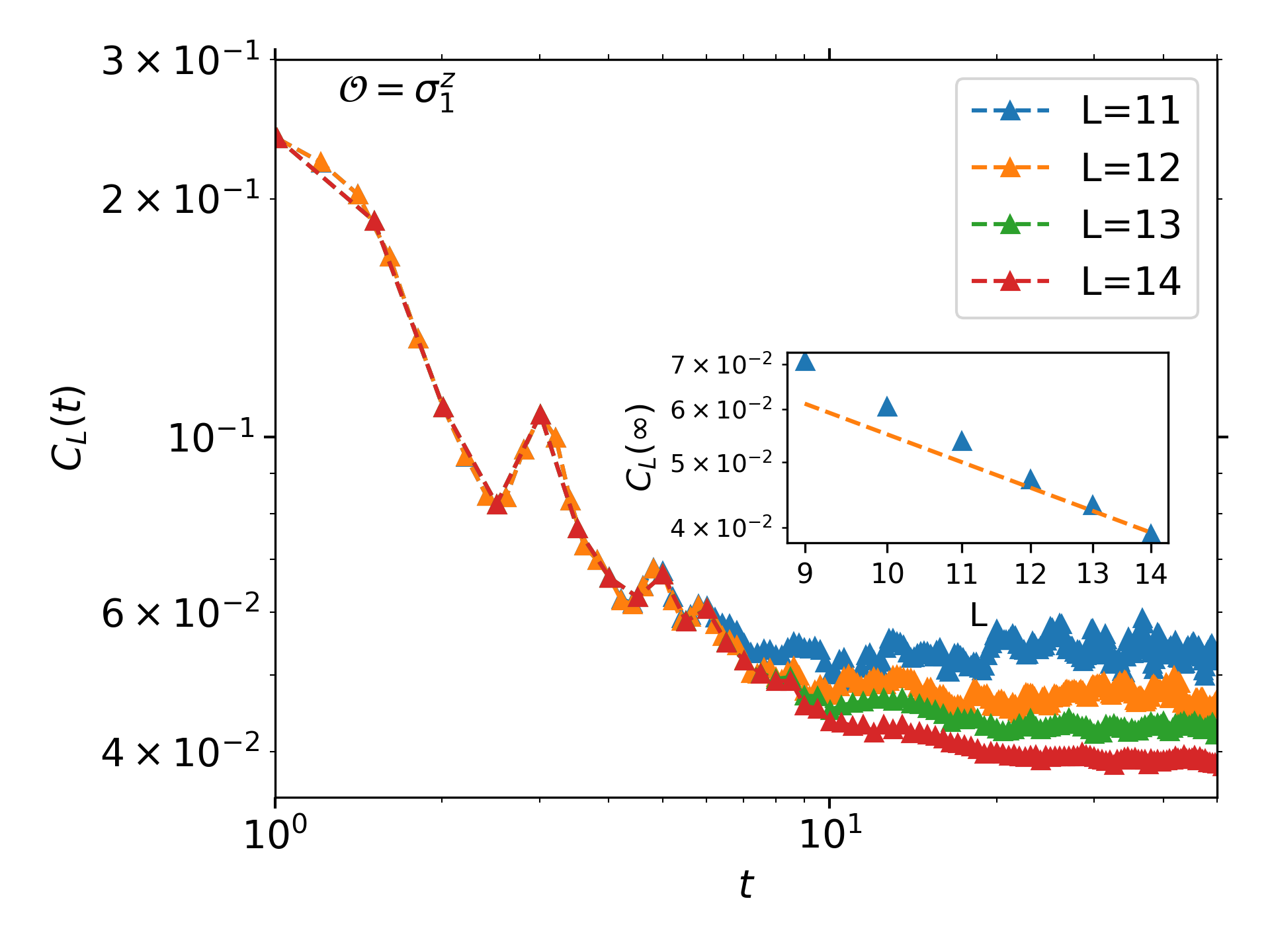}
\includegraphics[width=0.7\columnwidth]{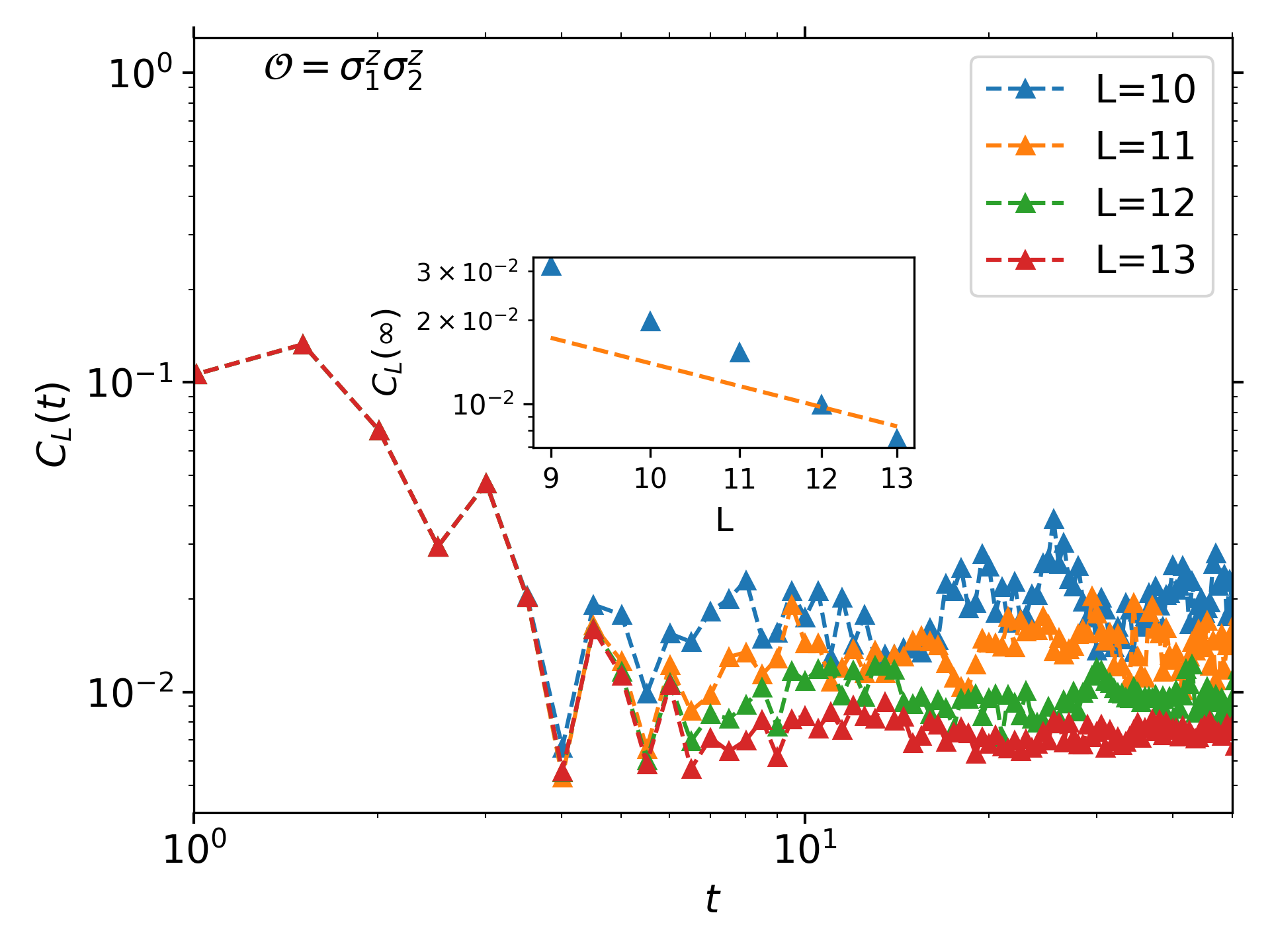}

\caption{Autocorrelation function of $\mathcal{O}=\sigma_1^z$ (top panel) and $\mathcal{O}=\sigma_1^z\sigma_2^z$ (bottom panel) using exact diagonalization. The insets are a fit of the plateau for different system sizes. While in the left inset the data for larger systems approach the scaling $L^{-1}$ of the orange dashed line, in the right inset the deviations from the reference line $L^{-2}$ are more evident.
}

\label{Fig:C_Z_plateau}
\end{figure}

In Fig.~\ref{Fig:Fn_comp}, we also apply the bi-exponential fitting of $F(n)\sim a_1 e^{- c_1 n}+ a_2 e^{- c_2 n}$ for the data with and without full orthogonalization. Remarkably, the fit function works very well in both cases and the fit parameters extracted from the Lanczos coefficients computed with full Gram–Schmidt orthogonalization are close to those obtained using the three-step algorithm in Eq.~\eqref{eq:3steps}, yielding a similar average rate $\bar{\Gamma}$.
Not only, thus, the data affected by the numerical error collapse according to the same qualitative function, they also display almost identical quantitative features.

Given the similarities of the data produced by the two algorithms, with and without full orthogonalization, we check the scaling rate of Eq.~(\textcolor{blue}{6}) in the main text for $m=2,3$ with the operators $\mathcal{O}=\sigma_1^z\sigma_2^z$ and $\mathcal{O}=\sigma_1^z\sigma_2^z\sigma_3^z$, respectively, using the SA in Eq.~\eqref{eq:3steps} in Fig.~\ref{fig:Fn_long}. This allows us to reach larger values of $n\sim 6000$. These data are out of reach if one considers the LA with full orthogonalization. We notice that a slow collapse of the data is visible only at values of $n$ which are larger than those considered in Fig.~\textcolor{blue}{3} of the main text for $\mathcal{O}=\sigma_1^z$. 
This supports our conjecture; 
the reason why the collapse was not visible for the operators $\sigma^z_1 \sigma^z_2$ and $\sigma^z_1 \sigma^z_2 \sigma^z_3$ using the LA with full orthogonalization (and was not presented in the main text) is likely that we could not compute the $b_n$ for large-enough $n$.
In summary, the results presented in this Appendix suggest that the standard Lanczos algorithm gives access to the universal features of the finite-size Lanczos coefficients $b_n$ no matter its numerical instability.

We conclude this Appendix by examining the origin of the imperfect data collapse in our plots of $-\log F(n)\cdot L^{m+1}/n (m=1,2)$
 shown in Fig.~\textcolor{blue}{3} of the main text and Fig.~\ref{fig:Fn_long}, respectively. Our results rely on the assumption that the plateaus $C_L(\infty)$ scale as $L^{-(m+1)}$. For the system sizes we have access to using exact diagonalization, we show in Fig.~\ref{Fig:C_Z_plateau} that the plateau of the autocorrelations of $\sigma^z_1$ (top panel) and $\sigma^z_1\sigma^z_2$ (bottom panel) decays faster than $L^{-1}$ or $L^{-2}$, respectively (especially in the latter case), due to finite size effects. This explains why the collapse of the data of the cumulative products is not perfect.
 We are convinced that if we could examine larger systems, we could observe the scalings predicted in the main text.

\end{appendix}

\bibliography{bibliography.bib}

\begin{thebibliography}{50}%
\makeatletter
\providecommand \@ifxundefined [1]{%
 \@ifx{#1\undefined}
}%
\providecommand \@ifnum [1]{%
 \ifnum #1\expandafter \@firstoftwo
 \else \expandafter \@secondoftwo
 \fi
}%
\providecommand \@ifx [1]{%
 \ifx #1\expandafter \@firstoftwo
 \else \expandafter \@secondoftwo
 \fi
}%
\providecommand \natexlab [1]{#1}%
\providecommand \enquote  [1]{``#1''}%
\providecommand \bibnamefont  [1]{#1}%
\providecommand \bibfnamefont [1]{#1}%
\providecommand \citenamefont [1]{#1}%
\providecommand \href@noop [0]{\@secondoftwo}%
\providecommand \href [0]{\begingroup \@sanitize@url \@href}%
\providecommand \@href[1]{\@@startlink{#1}\@@href}%
\providecommand \@@href[1]{\endgroup#1\@@endlink}%
\providecommand \@sanitize@url [0]{\catcode `\\12\catcode `\$12\catcode
  `\&12\catcode `\#12\catcode `\^12\catcode `\_12\catcode `\%12\relax}%
\providecommand \@@startlink[1]{}%
\providecommand \@@endlink[0]{}%
\providecommand \url  [0]{\begingroup\@sanitize@url \@url }%
\providecommand \@url [1]{\endgroup\@href {#1}{\urlprefix }}%
\providecommand \urlprefix  [0]{URL }%
\providecommand \Eprint [0]{\href }%
\providecommand \doibase [0]{https://doi.org/}%
\providecommand \selectlanguage [0]{\@gobble}%
\providecommand \bibinfo  [0]{\@secondoftwo}%
\providecommand \bibfield  [0]{\@secondoftwo}%
\providecommand \translation [1]{[#1]}%
\providecommand \BibitemOpen [0]{}%
\providecommand \bibitemStop [0]{}%
\providecommand \bibitemNoStop [0]{.\EOS\space}%
\providecommand \EOS [0]{\spacefactor3000\relax}%
\providecommand \BibitemShut  [1]{\csname bibitem#1\endcsname}%
\let\auto@bib@innerbib\@empty
\bibitem [{\citenamefont {Deutsch}(1991)}]{Deutsch-91}%
  \BibitemOpen
  \bibfield  {author} {\bibinfo {author} {\bibfnamefont {J.~M.}\ \bibnamefont
  {Deutsch}},\ }\bibfield  {title} {\bibinfo {title} {Quantum statistical
  mechanics in a closed system},\ }\href
  {https://doi.org/10.1103/PhysRevA.43.2046} {\bibfield  {journal} {\bibinfo
  {journal} {Phys. Rev. A}\ }\textbf {\bibinfo {volume} {43}},\ \bibinfo
  {pages} {2046} (\bibinfo {year} {1991})}\BibitemShut {NoStop}%
\bibitem [{\citenamefont {Srednicki}(1999)}]{Srednicki-99}%
  \BibitemOpen
  \bibfield  {author} {\bibinfo {author} {\bibfnamefont {M.}~\bibnamefont
  {Srednicki}},\ }\bibfield  {title} {\bibinfo {title} {The approach to thermal
  equilibrium in quantized chaotic systems},\ }\href
  {https://iopscience.iop.org/article/10.1088/0305-4470/32/7/007} {\bibfield
  {journal} {\bibinfo  {journal} {Journal of Physics A: Mathematical and
  General}\ }\textbf {\bibinfo {volume} {32}},\ \bibinfo {pages} {1163}
  (\bibinfo {year} {1999})}\BibitemShut {NoStop}%
\bibitem [{\citenamefont {Kadanoff}\ and\ \citenamefont
  {Martin}(1963)}]{km-63}%
  \BibitemOpen
  \bibfield  {author} {\bibinfo {author} {\bibfnamefont {L.~P.}\ \bibnamefont
  {Kadanoff}}\ and\ \bibinfo {author} {\bibfnamefont {P.~C.}\ \bibnamefont
  {Martin}},\ }\bibfield  {title} {\bibinfo {title} {Hydrodynamic equations and
  correlation functions},\ }\href {https://doi.org/10.1016/aphy.2000.6023}
  {\bibfield  {journal} {\bibinfo  {journal} {Annals of Physics}\ }\textbf
  {\bibinfo {volume} {24}},\ \bibinfo {pages} {419} (\bibinfo {year}
  {1963})}\BibitemShut {NoStop}%
\bibitem [{\citenamefont {Spohn}(2012)}]{Spohn-12}%
  \BibitemOpen
  \bibfield  {author} {\bibinfo {author} {\bibfnamefont {H.}~\bibnamefont
  {Spohn}},\ }\href@noop {} {\emph {\bibinfo {title} {Large scale dynamics of
  interacting particles}}}\ (\bibinfo  {publisher} {Springer Science \&
  Business Media},\ \bibinfo {year} {2012})\BibitemShut {NoStop}%
\bibitem [{\citenamefont {Castro-Alvaredo}\ \emph {et~al.}(2016)\citenamefont
  {Castro-Alvaredo}, \citenamefont {Doyon},\ and\ \citenamefont
  {Yoshimura}}]{Castro-16}%
  \BibitemOpen
  \bibfield  {author} {\bibinfo {author} {\bibfnamefont {O.~A.}\ \bibnamefont
  {Castro-Alvaredo}}, \bibinfo {author} {\bibfnamefont {B.}~\bibnamefont
  {Doyon}},\ and\ \bibinfo {author} {\bibfnamefont {T.}~\bibnamefont
  {Yoshimura}},\ }\bibfield  {title} {\bibinfo {title} {Emergent hydrodynamics
  in integrable quantum systems out of equilibrium},\ }\href
  {https://doi.org/10.1103/PhysRevX.6.041065} {\bibfield  {journal} {\bibinfo
  {journal} {Phys. Rev. X}\ }\textbf {\bibinfo {volume} {6}},\ \bibinfo {pages}
  {041065} (\bibinfo {year} {2016})}\BibitemShut {NoStop}%
\bibitem [{\citenamefont {Bertini}\ \emph {et~al.}(2016)\citenamefont
  {Bertini}, \citenamefont {Collura}, \citenamefont {De~Nardis},\ and\
  \citenamefont {Fagotti}}]{Bertini-16}%
  \BibitemOpen
  \bibfield  {author} {\bibinfo {author} {\bibfnamefont {B.}~\bibnamefont
  {Bertini}}, \bibinfo {author} {\bibfnamefont {M.}~\bibnamefont {Collura}},
  \bibinfo {author} {\bibfnamefont {J.}~\bibnamefont {De~Nardis}},\ and\
  \bibinfo {author} {\bibfnamefont {M.}~\bibnamefont {Fagotti}},\ }\bibfield
  {title} {\bibinfo {title} {{Transport in Out-of-Equilibrium XXZ Chains: Exact
  Profiles of Charges and Currents}},\ }\href
  {https://doi.org/10.1103/PhysRevLett.117.207201} {\bibfield  {journal}
  {\bibinfo  {journal} {Phys. Rev. Lett.}\ }\textbf {\bibinfo {volume} {117}},\
  \bibinfo {pages} {207201} (\bibinfo {year} {2016})}\BibitemShut {NoStop}%
\bibitem [{\citenamefont {Cirac}\ \emph {et~al.}(2021)\citenamefont {Cirac},
  \citenamefont {P\'erez-Garc\'{\i}a}, \citenamefont {Schuch},\ and\
  \citenamefont {Verstraete}}]{Cirac_2021}%
  \BibitemOpen
  \bibfield  {author} {\bibinfo {author} {\bibfnamefont {J.~I.}\ \bibnamefont
  {Cirac}}, \bibinfo {author} {\bibfnamefont {D.}~\bibnamefont
  {P\'erez-Garc\'{\i}a}}, \bibinfo {author} {\bibfnamefont {N.}~\bibnamefont
  {Schuch}},\ and\ \bibinfo {author} {\bibfnamefont {F.}~\bibnamefont
  {Verstraete}},\ }\bibfield  {title} {\bibinfo {title} {Matrix product states
  and projected entangled pair states: Concepts, symmetries, theorems},\ }\href
  {https://doi.org/10.1103/RevModPhys.93.045003} {\bibfield  {journal}
  {\bibinfo  {journal} {Rev. Mod. Phys.}\ }\textbf {\bibinfo {volume} {93}},\
  \bibinfo {pages} {045003} (\bibinfo {year} {2021})}\BibitemShut {NoStop}%
\bibitem [{\citenamefont {Rakovszky}\ \emph {et~al.}(2022)\citenamefont
  {Rakovszky}, \citenamefont {von Keyserlingk},\ and\ \citenamefont
  {Pollmann}}]{Rakovszky_2022}%
  \BibitemOpen
  \bibfield  {author} {\bibinfo {author} {\bibfnamefont {T.}~\bibnamefont
  {Rakovszky}}, \bibinfo {author} {\bibfnamefont {C.~W.}\ \bibnamefont {von
  Keyserlingk}},\ and\ \bibinfo {author} {\bibfnamefont {F.}~\bibnamefont
  {Pollmann}},\ }\bibfield  {title} {\bibinfo {title} {Dissipation-assisted
  operator evolution method for capturing hydrodynamic transport},\ }\href
  {https://doi.org/10.1103/PhysRevB.105.075131} {\bibfield  {journal} {\bibinfo
   {journal} {Phys. Rev. B}\ }\textbf {\bibinfo {volume} {105}},\ \bibinfo
  {pages} {075131} (\bibinfo {year} {2022})}\BibitemShut {NoStop}%
\bibitem [{\citenamefont {Viswanath}\ and\ \citenamefont
  {M\"uller}(1994)}]{Recursion-1994}%
  \BibitemOpen
  \bibfield  {author} {\bibinfo {author} {\bibfnamefont {V.~S.}\ \bibnamefont
  {Viswanath}}\ and\ \bibinfo {author} {\bibfnamefont {G.}~\bibnamefont
  {M\"uller}},\ }\href@noop {} {\emph {\bibinfo {title} {The Recursion Method:
  Application to Many-Body Dynamics}}}\ (\bibinfo  {publisher} {Springer
  Science \& Business Media},\ \bibinfo {year} {1994})\BibitemShut {NoStop}%
\bibitem [{\citenamefont {Nandy}\ \emph {et~al.}(2025)\citenamefont {Nandy},
  \citenamefont {Matsoukas-Roubeas}, \citenamefont {Martínez-Azcona},
  \citenamefont {Dymarsky},\ and\ \citenamefont {del Campo}}]{Nandy-2025}%
  \BibitemOpen
  \bibfield  {author} {\bibinfo {author} {\bibfnamefont {P.}~\bibnamefont
  {Nandy}}, \bibinfo {author} {\bibfnamefont {A.~S.}\ \bibnamefont
  {Matsoukas-Roubeas}}, \bibinfo {author} {\bibfnamefont {P.}~\bibnamefont
  {Martínez-Azcona}}, \bibinfo {author} {\bibfnamefont {A.}~\bibnamefont
  {Dymarsky}},\ and\ \bibinfo {author} {\bibfnamefont {A.}~\bibnamefont {del
  Campo}},\ }\bibfield  {title} {\bibinfo {title} {Quantum dynamics in krylov
  space: Methods and applications},\ }\href
  {https://doi.org/10.1016/j.physrep.2025.05.001} {\bibfield  {journal}
  {\bibinfo  {journal} {Physics Reports}\ }\textbf {\bibinfo {volume}
  {1125–1128}},\ \bibinfo {pages} {1–82} (\bibinfo {year}
  {2025})}\BibitemShut {NoStop}%
\bibitem [{\citenamefont {Parker}\ \emph {et~al.}(2019)\citenamefont {Parker},
  \citenamefont {Cao}, \citenamefont {Avdoshkin}, \citenamefont {Scaffidi},\
  and\ \citenamefont {Altman}}]{Parker-19}%
  \BibitemOpen
  \bibfield  {author} {\bibinfo {author} {\bibfnamefont {D.~E.}\ \bibnamefont
  {Parker}}, \bibinfo {author} {\bibfnamefont {X.}~\bibnamefont {Cao}},
  \bibinfo {author} {\bibfnamefont {A.}~\bibnamefont {Avdoshkin}}, \bibinfo
  {author} {\bibfnamefont {T.}~\bibnamefont {Scaffidi}},\ and\ \bibinfo
  {author} {\bibfnamefont {E.}~\bibnamefont {Altman}},\ }\bibfield  {title}
  {\bibinfo {title} {A universal operator growth hypothesis},\ }\href
  {https://doi.org/10.1103/PhysRevX.9.041017} {\bibfield  {journal} {\bibinfo
  {journal} {Phys. Rev. X}\ }\textbf {\bibinfo {volume} {9}},\ \bibinfo {pages}
  {041017} (\bibinfo {year} {2019})}\BibitemShut {NoStop}%
\bibitem [{\citenamefont {Yates}\ \emph
  {et~al.}(2020{\natexlab{a}})\citenamefont {Yates}, \citenamefont {Abanov},\
  and\ \citenamefont {Mitra}}]{Yates2020PRL}%
  \BibitemOpen
  \bibfield  {author} {\bibinfo {author} {\bibfnamefont {D.~J.}\ \bibnamefont
  {Yates}}, \bibinfo {author} {\bibfnamefont {A.~G.}\ \bibnamefont {Abanov}},\
  and\ \bibinfo {author} {\bibfnamefont {A.}~\bibnamefont {Mitra}},\ }\bibfield
   {title} {\bibinfo {title} {Lifetime of almost strong edge-mode operators in
  one-dimensional, interacting, symmetry protected topological phases},\ }\href
  {https://doi.org/10.1103/PhysRevLett.124.206803} {\bibfield  {journal}
  {\bibinfo  {journal} {Phys. Rev. Lett.}\ }\textbf {\bibinfo {volume} {124}},\
  \bibinfo {pages} {206803} (\bibinfo {year} {2020}{\natexlab{a}})}\BibitemShut
  {NoStop}%
\bibitem [{\citenamefont {Yates}\ \emph
  {et~al.}(2020{\natexlab{b}})\citenamefont {Yates}, \citenamefont {Abanov},\
  and\ \citenamefont {Mitra}}]{yam-20}%
  \BibitemOpen
  \bibfield  {author} {\bibinfo {author} {\bibfnamefont {D.~J.}\ \bibnamefont
  {Yates}}, \bibinfo {author} {\bibfnamefont {A.~G.}\ \bibnamefont {Abanov}},\
  and\ \bibinfo {author} {\bibfnamefont {A.}~\bibnamefont {Mitra}},\ }\bibfield
   {title} {\bibinfo {title} {Dynamics of almost strong edge modes in spin
  chains away from integrability},\ }\href
  {https://doi.org/10.1103/PhysRevB.102.195419} {\bibfield  {journal} {\bibinfo
   {journal} {Phys. Rev. B}\ }\textbf {\bibinfo {volume} {102}},\ \bibinfo
  {pages} {195419} (\bibinfo {year} {2020}{\natexlab{b}})}\BibitemShut
  {NoStop}%
\bibitem [{\citenamefont {Yates}\ and\ \citenamefont
  {Mitra}(2021)}]{yates2021PRB}%
  \BibitemOpen
  \bibfield  {author} {\bibinfo {author} {\bibfnamefont {D.~J.}\ \bibnamefont
  {Yates}}\ and\ \bibinfo {author} {\bibfnamefont {A.}~\bibnamefont {Mitra}},\
  }\bibfield  {title} {\bibinfo {title} {{Strong and almost strong modes of
  Floquet spin chains in Krylov subspaces}},\ }\href
  {https://doi.org/10.1103/PhysRevB.104.195121} {\bibfield  {journal} {\bibinfo
   {journal} {Phys. Rev. B}\ }\textbf {\bibinfo {volume} {104}},\ \bibinfo
  {pages} {195121} (\bibinfo {year} {2021})}\BibitemShut {NoStop}%
\bibitem [{\citenamefont {Yeh}\ \emph {et~al.}(2023)\citenamefont {Yeh},
  \citenamefont {Cardoso}, \citenamefont {Korneev}, \citenamefont {Sels},
  \citenamefont {Abanov},\ and\ \citenamefont {Mitra}}]{Yeh2023}%
  \BibitemOpen
  \bibfield  {author} {\bibinfo {author} {\bibfnamefont {H.-C.}\ \bibnamefont
  {Yeh}}, \bibinfo {author} {\bibfnamefont {G.}~\bibnamefont {Cardoso}},
  \bibinfo {author} {\bibfnamefont {L.}~\bibnamefont {Korneev}}, \bibinfo
  {author} {\bibfnamefont {D.}~\bibnamefont {Sels}}, \bibinfo {author}
  {\bibfnamefont {A.~G.}\ \bibnamefont {Abanov}},\ and\ \bibinfo {author}
  {\bibfnamefont {A.}~\bibnamefont {Mitra}},\ }\bibfield  {title} {\bibinfo
  {title} {Slowly decaying zero mode in a weakly nonintegrable boundary
  impurity model},\ }\href {https://doi.org/10.1103/PhysRevB.108.165143}
  {\bibfield  {journal} {\bibinfo  {journal} {Phys. Rev. B}\ }\textbf {\bibinfo
  {volume} {108}},\ \bibinfo {pages} {165143} (\bibinfo {year}
  {2023})}\BibitemShut {NoStop}%
\bibitem [{\citenamefont {Tausendpfund}\ \emph {et~al.}(2025)\citenamefont
  {Tausendpfund}, \citenamefont {Mitra},\ and\ \citenamefont {Rizzi}}]{tmr-25}%
  \BibitemOpen
  \bibfield  {author} {\bibinfo {author} {\bibfnamefont {N.}~\bibnamefont
  {Tausendpfund}}, \bibinfo {author} {\bibfnamefont {A.}~\bibnamefont
  {Mitra}},\ and\ \bibinfo {author} {\bibfnamefont {M.}~\bibnamefont {Rizzi}},\
  }\href {https://arxiv.org/abs/2501.11121} {\bibinfo {title} {Almost strong
  zero modes at finite temperature}} (\bibinfo {year} {2025}),\ \Eprint
  {https://arxiv.org/abs/2501.11121} {arXiv:2501.11121 [cond-mat.str-el]}
  \BibitemShut {NoStop}%
\bibitem [{\citenamefont {Yeh}\ and\ \citenamefont {Mitra}(2024)}]{yeh2024PRB}%
  \BibitemOpen
  \bibfield  {author} {\bibinfo {author} {\bibfnamefont {H.-C.}\ \bibnamefont
  {Yeh}}\ and\ \bibinfo {author} {\bibfnamefont {A.}~\bibnamefont {Mitra}},\
  }\bibfield  {title} {\bibinfo {title} {{Universal model of Floquet operator
  Krylov space}},\ }\href {https://doi.org/10.1103/PhysRevB.110.155109}
  {\bibfield  {journal} {\bibinfo  {journal} {Phys. Rev. B}\ }\textbf {\bibinfo
  {volume} {110}},\ \bibinfo {pages} {155109} (\bibinfo {year}
  {2024})}\BibitemShut {NoStop}%
\bibitem [{\citenamefont {Yeh}\ and\ \citenamefont {Mitra}(2025)}]{yeh2025}%
  \BibitemOpen
  \bibfield  {author} {\bibinfo {author} {\bibfnamefont {H.-C.}\ \bibnamefont
  {Yeh}}\ and\ \bibinfo {author} {\bibfnamefont {A.}~\bibnamefont {Mitra}},\
  }\bibfield  {title} {\bibinfo {title} {{Moment method and continued fraction
  expansion in Floquet operator Krylov space}},\ }\href
  {https://doi.org/10.1103/PhysRevB.111.125103} {\bibfield  {journal} {\bibinfo
   {journal} {Phys. Rev. B}\ }\textbf {\bibinfo {volume} {111}},\ \bibinfo
  {pages} {125103} (\bibinfo {year} {2025})}\BibitemShut {NoStop}%
\bibitem [{\citenamefont {Yates}\ and\ \citenamefont
  {Mitra}(2022)}]{yates2022comm}%
  \BibitemOpen
  \bibfield  {author} {\bibinfo {author} {\bibfnamefont {A.~G.}\ \bibnamefont
  {Yates}, \bibfnamefont {Daniel J.~Abanov}}\ and\ \bibinfo {author}
  {\bibfnamefont {A.}~\bibnamefont {Mitra}},\ }\bibfield  {title} {\bibinfo
  {title} {{Long-lived period-doubled edge modes of interacting and
  disorder-free Floquet spin chains}},\ }\bibfield  {journal} {\bibinfo
  {journal} {Commun Phys}\ }\textbf {\bibinfo {volume} {5}},\ \href
  {https://doi.org/10.1038/s42005-022-00818-1} {10.1038/s42005-022-00818-1}
  (\bibinfo {year} {2022})\BibitemShut {NoStop}%
\bibitem [{\citenamefont {Suchsland}\ \emph {et~al.}(2025)\citenamefont
  {Suchsland}, \citenamefont {Moessner},\ and\ \citenamefont
  {Claeys}}]{Suchsland2025}%
  \BibitemOpen
  \bibfield  {author} {\bibinfo {author} {\bibfnamefont {P.}~\bibnamefont
  {Suchsland}}, \bibinfo {author} {\bibfnamefont {R.}~\bibnamefont
  {Moessner}},\ and\ \bibinfo {author} {\bibfnamefont {P.~W.}\ \bibnamefont
  {Claeys}},\ }\bibfield  {title} {\bibinfo {title} {{Krylov complexity and
  Trotter transitions in unitary circuit dynamics}},\ }\href
  {https://doi.org/10.1103/PhysRevB.111.014309} {\bibfield  {journal} {\bibinfo
   {journal} {Phys. Rev. B}\ }\textbf {\bibinfo {volume} {111}},\ \bibinfo
  {pages} {014309} (\bibinfo {year} {2025})}\BibitemShut {NoStop}%
\bibitem [{\citenamefont {Kolganov}\ and\ \citenamefont
  {Trunin}(2025)}]{Kolganov2025}%
  \BibitemOpen
  \bibfield  {author} {\bibinfo {author} {\bibfnamefont {N.}~\bibnamefont
  {Kolganov}}\ and\ \bibinfo {author} {\bibfnamefont {D.~A.}\ \bibnamefont
  {Trunin}},\ }\bibfield  {title} {\bibinfo {title} {Streamlined krylov
  construction and classification of ergodic floquet systems},\ }\href
  {https://doi.org/10.1103/PhysRevE.111.L052202} {\bibfield  {journal}
  {\bibinfo  {journal} {Phys. Rev. E}\ }\textbf {\bibinfo {volume} {111}},\
  \bibinfo {pages} {L052202} (\bibinfo {year} {2025})}\BibitemShut {NoStop}%
\bibitem [{\citenamefont {{Rabinovici, E. and Sánchez-Garrido, A. and Shir,
  Ruth and Sonner, J.}}(2023)}]{Rabinovici2023}%
  \BibitemOpen
  \bibfield  {author} {\bibinfo {author} {\bibnamefont {{Rabinovici, E. and
  Sánchez-Garrido, A. and Shir, Ruth and Sonner, J.}}},\ }\bibfield  {title}
  {\bibinfo {title} {A bulk manifestation of krylov complexity},\ }\href
  {https://doi.org/10.1007/JHEP08(2023)213} {\bibfield  {journal} {\bibinfo
  {journal} {Journal of High Energy Physics}\ }\textbf {\bibinfo {volume}
  {2023}} (\bibinfo {year} {2023})}\BibitemShut {NoStop}%
\bibitem [{\citenamefont {Balasubramanian}\ \emph {et~al.}(2024)\citenamefont
  {Balasubramanian}, \citenamefont {Magan}, \citenamefont {Nandi},\ and\
  \citenamefont {Wu}}]{Balasubramanian2024}%
  \BibitemOpen
  \bibfield  {author} {\bibinfo {author} {\bibfnamefont {V.}~\bibnamefont
  {Balasubramanian}}, \bibinfo {author} {\bibfnamefont {J.~M.}\ \bibnamefont
  {Magan}}, \bibinfo {author} {\bibfnamefont {P.}~\bibnamefont {Nandi}},\ and\
  \bibinfo {author} {\bibfnamefont {Q.}~\bibnamefont {Wu}},\ }\href
  {https://arxiv.org/abs/2412.02038} {\bibinfo {title} {Spread complexity and
  the saturation of wormhole size}} (\bibinfo {year} {2024}),\ \Eprint
  {https://arxiv.org/abs/2412.02038} {arXiv:2412.02038 [hep-th]} \BibitemShut
  {NoStop}%
\bibitem [{\citenamefont {Caputa}\ \emph {et~al.}(2024)\citenamefont {Caputa},
  \citenamefont {Chen}, \citenamefont {McDonald}, \citenamefont {Simón},\ and\
  \citenamefont {Strittmatter}}]{Caputa2024}%
  \BibitemOpen
  \bibfield  {author} {\bibinfo {author} {\bibfnamefont {P.}~\bibnamefont
  {Caputa}}, \bibinfo {author} {\bibfnamefont {B.}~\bibnamefont {Chen}},
  \bibinfo {author} {\bibfnamefont {R.~W.}\ \bibnamefont {McDonald}}, \bibinfo
  {author} {\bibfnamefont {J.}~\bibnamefont {Simón}},\ and\ \bibinfo {author}
  {\bibfnamefont {B.}~\bibnamefont {Strittmatter}},\ }\href
  {https://arxiv.org/abs/2410.23334} {\bibinfo {title} {Spread complexity rate
  as proper momentum}} (\bibinfo {year} {2024}),\ \Eprint
  {https://arxiv.org/abs/2410.23334} {arXiv:2410.23334 [hep-th]} \BibitemShut
  {NoStop}%
\bibitem [{\citenamefont {Miyaji}\ \emph {et~al.}(2025)\citenamefont {Miyaji},
  \citenamefont {Ruan}, \citenamefont {Shibuya},\ and\ \citenamefont
  {Yano}}]{Miyaji2025}%
  \BibitemOpen
  \bibfield  {author} {\bibinfo {author} {\bibfnamefont {M.}~\bibnamefont
  {Miyaji}}, \bibinfo {author} {\bibfnamefont {S.-M.}\ \bibnamefont {Ruan}},
  \bibinfo {author} {\bibfnamefont {S.}~\bibnamefont {Shibuya}},\ and\ \bibinfo
  {author} {\bibfnamefont {K.}~\bibnamefont {Yano}},\ }\href
  {https://arxiv.org/abs/2502.12266} {\bibinfo {title} {{Non-perturbative
  Overlaps in JT Gravity: From Spectral Form Factor to Generating Functions of
  Complexity}}} (\bibinfo {year} {2025}),\ \Eprint
  {https://arxiv.org/abs/2502.12266} {arXiv:2502.12266 [hep-th]} \BibitemShut
  {NoStop}%
\bibitem [{\citenamefont {Caputa}\ and\ \citenamefont
  {Giulio}(2025)}]{Caputa2025}%
  \BibitemOpen
  \bibfield  {author} {\bibinfo {author} {\bibfnamefont {P.}~\bibnamefont
  {Caputa}}\ and\ \bibinfo {author} {\bibfnamefont {G.~D.}\ \bibnamefont
  {Giulio}},\ }\href {https://arxiv.org/abs/2502.19485} {\bibinfo {title}
  {{Local Quenches from a Krylov Perspective}}} (\bibinfo {year} {2025}),\
  \Eprint {https://arxiv.org/abs/2502.19485} {arXiv:2502.19485 [hep-th]}
  \BibitemShut {NoStop}%
\bibitem [{\citenamefont {Avdoshkin}\ \emph {et~al.}(2024)\citenamefont
  {Avdoshkin}, \citenamefont {Dymarsky},\ and\ \citenamefont
  {Smolkin}}]{Avdoshkin2022}%
  \BibitemOpen
  \bibfield  {author} {\bibinfo {author} {\bibfnamefont {A.}~\bibnamefont
  {Avdoshkin}}, \bibinfo {author} {\bibfnamefont {A.}~\bibnamefont
  {Dymarsky}},\ and\ \bibinfo {author} {\bibfnamefont {M.}~\bibnamefont
  {Smolkin}},\ }\bibfield  {title} {\bibinfo {title} {{Krylov complexity in
  quantum field theory, and beyond}},\ }\href
  {https://doi.org/10.1007/JHEP06(2024)066} {\bibfield  {journal} {\bibinfo
  {journal} {Journal of High Energy Physics}\ }\textbf {\bibinfo {volume}
  {2024}} (\bibinfo {year} {2024})}\BibitemShut {NoStop}%
\bibitem [{\citenamefont {Dymarsky}\ and\ \citenamefont
  {Smolkin}(2021)}]{Dymarsky2021}%
  \BibitemOpen
  \bibfield  {author} {\bibinfo {author} {\bibfnamefont {A.}~\bibnamefont
  {Dymarsky}}\ and\ \bibinfo {author} {\bibfnamefont {M.}~\bibnamefont
  {Smolkin}},\ }\bibfield  {title} {\bibinfo {title} {{{Krylov complexity in
  conformal field theory}}},\ }\href
  {https://doi.org/10.1103/PhysRevD.104.L081702} {\bibfield  {journal}
  {\bibinfo  {journal} {Phys. Rev. D}\ }\textbf {\bibinfo {volume} {104}},\
  \bibinfo {pages} {L081702} (\bibinfo {year} {2021})},\ \Eprint
  {https://arxiv.org/abs/2104.09514} {arXiv:2104.09514 [hep-th]} \BibitemShut
  {NoStop}%
\bibitem [{\citenamefont {Bhattacharjee}\ \emph {et~al.}(2023)\citenamefont
  {Bhattacharjee}, \citenamefont {Cao}, \citenamefont {Nandy},\ and\
  \citenamefont {Pathak}}]{Bhattacharjee2022operator}%
  \BibitemOpen
  \bibfield  {author} {\bibinfo {author} {\bibfnamefont {B.}~\bibnamefont
  {Bhattacharjee}}, \bibinfo {author} {\bibfnamefont {X.}~\bibnamefont {Cao}},
  \bibinfo {author} {\bibfnamefont {P.}~\bibnamefont {Nandy}},\ and\ \bibinfo
  {author} {\bibfnamefont {T.}~\bibnamefont {Pathak}},\ }\bibfield  {title}
  {\bibinfo {title} {{Operator growth in open quantum systems: lessons from the
  dissipative SYK}},\ }\href {https://doi.org/10.1007/JHEP03(2023)054}
  {\bibfield  {journal} {\bibinfo  {journal} {Journal of High Energy Physics}\
  }\textbf {\bibinfo {volume} {2023}} (\bibinfo {year} {2023})}\BibitemShut
  {NoStop}%
\bibitem [{\citenamefont {Balasubramanian}\ \emph {et~al.}(2023)\citenamefont
  {Balasubramanian}, \citenamefont {Magan},\ and\ \citenamefont
  {Wu}}]{Balasubramanian2022dnj}%
  \BibitemOpen
  \bibfield  {author} {\bibinfo {author} {\bibfnamefont {V.}~\bibnamefont
  {Balasubramanian}}, \bibinfo {author} {\bibfnamefont {J.~M.}\ \bibnamefont
  {Magan}},\ and\ \bibinfo {author} {\bibfnamefont {Q.}~\bibnamefont {Wu}},\
  }\bibfield  {title} {\bibinfo {title} {{Tridiagonalizing random matrices}},\
  }\href {https://doi.org/10.1103/PhysRevD.107.126001} {\bibfield  {journal}
  {\bibinfo  {journal} {Phys. Rev. D}\ }\textbf {\bibinfo {volume} {107}},\
  \bibinfo {pages} {126001} (\bibinfo {year} {2023})},\ \Eprint
  {https://arxiv.org/abs/2208.08452} {arXiv:2208.08452 [hep-th]} \BibitemShut
  {NoStop}%
\bibitem [{\citenamefont {Balasubramanian}\ \emph {et~al.}(2025)\citenamefont
  {Balasubramanian}, \citenamefont {Das}, \citenamefont {Erdmenger},\ and\
  \citenamefont {Xian}}]{Balasubramanian2024ghv}%
  \BibitemOpen
  \bibfield  {author} {\bibinfo {author} {\bibfnamefont {V.}~\bibnamefont
  {Balasubramanian}}, \bibinfo {author} {\bibfnamefont {R.~N.}\ \bibnamefont
  {Das}}, \bibinfo {author} {\bibfnamefont {J.}~\bibnamefont {Erdmenger}},\
  and\ \bibinfo {author} {\bibfnamefont {Z.-Y.}\ \bibnamefont {Xian}},\
  }\bibfield  {title} {\bibinfo {title} {{Chaos and integrability in triangular
  billiards}},\ }\href {https://doi.org/10.1088/1742-5468/adba41} {\bibfield
  {journal} {\bibinfo  {journal} {J. Stat. Mech.}\ }\textbf {\bibinfo {volume}
  {2025}},\ \bibinfo {pages} {033202} (\bibinfo {year} {2025})},\ \Eprint
  {https://arxiv.org/abs/2407.11114} {arXiv:2407.11114 [hep-th]} \BibitemShut
  {NoStop}%
\bibitem [{\citenamefont {Erdmenger}\ \emph {et~al.}(2023)\citenamefont
  {Erdmenger}, \citenamefont {Jian},\ and\ \citenamefont
  {Xian}}]{Erdmenger2023}%
  \BibitemOpen
  \bibfield  {author} {\bibinfo {author} {\bibfnamefont {J.}~\bibnamefont
  {Erdmenger}}, \bibinfo {author} {\bibfnamefont {S.}~\bibnamefont {Jian}},\
  and\ \bibinfo {author} {\bibfnamefont {Z.-Y.}\ \bibnamefont {Xian}},\
  }\bibfield  {title} {\bibinfo {title} {{Universal chaotic dynamics from
  Krylov space}},\ }\href {https://doi.org/10.1007/JHEP08(2023)176} {\bibfield
  {journal} {\bibinfo  {journal} {Journal of High Energy Physics}\ }\textbf
  {\bibinfo {volume} {2023}} (\bibinfo {year} {2023})}\BibitemShut {NoStop}%
\bibitem [{\citenamefont {Rabinovici}\ \emph {et~al.}(2022)\citenamefont
  {Rabinovici}, \citenamefont {Sánchez-Garrido}, \citenamefont {Shir},\ and\
  \citenamefont {Sonner}}]{Rabinovici2022}%
  \BibitemOpen
  \bibfield  {author} {\bibinfo {author} {\bibfnamefont {E.}~\bibnamefont
  {Rabinovici}}, \bibinfo {author} {\bibfnamefont {A.}~\bibnamefont
  {Sánchez-Garrido}}, \bibinfo {author} {\bibfnamefont {R.}~\bibnamefont
  {Shir}},\ and\ \bibinfo {author} {\bibfnamefont {J.}~\bibnamefont {Sonner}},\
  }\bibfield  {title} {\bibinfo {title} {{Krylov complexity from integrability
  to chaos}},\ }\href {https://doi.org/10.1007/JHEP07(2022)151} {\bibfield
  {journal} {\bibinfo  {journal} {Journal of High Energy Physics}\ }\textbf
  {\bibinfo {volume} {2022}} (\bibinfo {year} {2022})}\BibitemShut {NoStop}%
\bibitem [{\citenamefont {Caputa}\ \emph {et~al.}(2025)\citenamefont {Caputa},
  \citenamefont {Giulio},\ and\ \citenamefont {Loc}}]{Caputa2025gro}%
  \BibitemOpen
  \bibfield  {author} {\bibinfo {author} {\bibfnamefont {P.}~\bibnamefont
  {Caputa}}, \bibinfo {author} {\bibfnamefont {G.~D.}\ \bibnamefont {Giulio}},\
  and\ \bibinfo {author} {\bibfnamefont {T.~Q.}\ \bibnamefont {Loc}},\ }\href
  {https://arxiv.org/abs/2507.02033} {\bibinfo {title} {{Growth of block
  diagonal operators and symmetry-resolved Krylov complexity}}} (\bibinfo
  {year} {2025}),\ \Eprint {https://arxiv.org/abs/2507.02033} {arXiv:2507.02033
  [hep-th]} \BibitemShut {NoStop}%
\bibitem [{\citenamefont {Bhattacharjee}\ \emph {et~al.}(2022)\citenamefont
  {Bhattacharjee}, \citenamefont {Cao}, \citenamefont {Nandy},\ and\
  \citenamefont {Pathak}}]{Bhattacharjee_2022}%
  \BibitemOpen
  \bibfield  {author} {\bibinfo {author} {\bibfnamefont {B.}~\bibnamefont
  {Bhattacharjee}}, \bibinfo {author} {\bibfnamefont {X.}~\bibnamefont {Cao}},
  \bibinfo {author} {\bibfnamefont {P.}~\bibnamefont {Nandy}},\ and\ \bibinfo
  {author} {\bibfnamefont {T.}~\bibnamefont {Pathak}},\ }\bibfield  {title}
  {\bibinfo {title} {Krylov complexity in saddle-dominated scrambling},\
  }\bibfield  {journal} {\bibinfo  {journal} {Journal of High Energy Physics}\
  }\textbf {\bibinfo {volume} {2022}},\ \href
  {https://doi.org/10.1007/jhep05(2022)174} {10.1007/jhep05(2022)174} (\bibinfo
  {year} {2022})\BibitemShut {NoStop}%
\bibitem [{\citenamefont {Cao}(2021)}]{Cao2021}%
  \BibitemOpen
  \bibfield  {author} {\bibinfo {author} {\bibfnamefont {X.}~\bibnamefont
  {Cao}},\ }\bibfield  {title} {\bibinfo {title} {A statistical mechanism for
  operator growth},\ }\href {https://doi.org/10.1088/1751-8121/abe77c}
  {\bibfield  {journal} {\bibinfo  {journal} {Journal of Physics A:
  Mathematical and Theoretical}\ }\textbf {\bibinfo {volume} {54}},\ \bibinfo
  {pages} {144001} (\bibinfo {year} {2021})}\BibitemShut {NoStop}%
\bibitem [{\citenamefont {Qi}\ \emph {et~al.}(2023)\citenamefont {Qi},
  \citenamefont {Scaffidi},\ and\ \citenamefont {Cao}}]{qi2023}%
  \BibitemOpen
  \bibfield  {author} {\bibinfo {author} {\bibfnamefont {Z.}~\bibnamefont
  {Qi}}, \bibinfo {author} {\bibfnamefont {T.}~\bibnamefont {Scaffidi}},\ and\
  \bibinfo {author} {\bibfnamefont {X.}~\bibnamefont {Cao}},\ }\bibfield
  {title} {\bibinfo {title} {Surprises in the deep hilbert space of all-to-all
  systems: From superexponential scrambling to slow entanglement growth},\
  }\href {https://doi.org/10.1103/PhysRevB.108.054301} {\bibfield  {journal}
  {\bibinfo  {journal} {Phys. Rev. B}\ }\textbf {\bibinfo {volume} {108}},\
  \bibinfo {pages} {054301} (\bibinfo {year} {2023})}\BibitemShut {NoStop}%
\bibitem [{\citenamefont {Uskov}\ and\ \citenamefont
  {Lychkovskiy}(2024)}]{Uskov2024}%
  \BibitemOpen
  \bibfield  {author} {\bibinfo {author} {\bibfnamefont {F.}~\bibnamefont
  {Uskov}}\ and\ \bibinfo {author} {\bibfnamefont {O.}~\bibnamefont
  {Lychkovskiy}},\ }\bibfield  {title} {\bibinfo {title} {Quantum dynamics in
  one and two dimensions via the recursion method},\ }\href
  {https://doi.org/10.1103/PhysRevB.109.L140301} {\bibfield  {journal}
  {\bibinfo  {journal} {Phys. Rev. B}\ }\textbf {\bibinfo {volume} {109}},\
  \bibinfo {pages} {L140301} (\bibinfo {year} {2024})}\BibitemShut {NoStop}%
\bibitem [{\citenamefont {Wang}\ \emph {et~al.}(2024)\citenamefont {Wang},
  \citenamefont {Lamann}, \citenamefont {Steinigeweg},\ and\ \citenamefont
  {Gemmer}}]{wang2024}%
  \BibitemOpen
  \bibfield  {author} {\bibinfo {author} {\bibfnamefont {J.}~\bibnamefont
  {Wang}}, \bibinfo {author} {\bibfnamefont {M.~H.}\ \bibnamefont {Lamann}},
  \bibinfo {author} {\bibfnamefont {R.}~\bibnamefont {Steinigeweg}},\ and\
  \bibinfo {author} {\bibfnamefont {J.}~\bibnamefont {Gemmer}},\ }\bibfield
  {title} {\bibinfo {title} {Diffusion constants from the recursion method},\
  }\href {https://doi.org/10.1103/PhysRevB.110.104413} {\bibfield  {journal}
  {\bibinfo  {journal} {Phys. Rev. B}\ }\textbf {\bibinfo {volume} {110}},\
  \bibinfo {pages} {104413} (\bibinfo {year} {2024})}\BibitemShut {NoStop}%
\bibitem [{\citenamefont {Yi-Thomas}\ \emph {et~al.}(2024)\citenamefont
  {Yi-Thomas}, \citenamefont {Ware}, \citenamefont {Sau},\ and\ \citenamefont
  {White}}]{yi2024}%
  \BibitemOpen
  \bibfield  {author} {\bibinfo {author} {\bibfnamefont {S.}~\bibnamefont
  {Yi-Thomas}}, \bibinfo {author} {\bibfnamefont {B.}~\bibnamefont {Ware}},
  \bibinfo {author} {\bibfnamefont {J.~D.}\ \bibnamefont {Sau}},\ and\ \bibinfo
  {author} {\bibfnamefont {C.~D.}\ \bibnamefont {White}},\ }\bibfield  {title}
  {\bibinfo {title} {Comparing numerical methods for hydrodynamics in a
  one-dimensional lattice spin model},\ }\href
  {https://doi.org/10.1103/PhysRevB.110.134308} {\bibfield  {journal} {\bibinfo
   {journal} {Phys. Rev. B}\ }\textbf {\bibinfo {volume} {110}},\ \bibinfo
  {pages} {134308} (\bibinfo {year} {2024})}\BibitemShut {NoStop}%
\bibitem [{\citenamefont {Loizeau}\ \emph
  {et~al.}(2025{\natexlab{a}})\citenamefont {Loizeau}, \citenamefont
  {Peacock},\ and\ \citenamefont {Sels}}]{Loizeau2025quantum}%
  \BibitemOpen
  \bibfield  {author} {\bibinfo {author} {\bibfnamefont {N.}~\bibnamefont
  {Loizeau}}, \bibinfo {author} {\bibfnamefont {J.~C.}\ \bibnamefont
  {Peacock}},\ and\ \bibinfo {author} {\bibfnamefont {D.}~\bibnamefont
  {Sels}},\ }\bibfield  {title} {\bibinfo {title} {Quantum many-body
  simulations with paulistrings.jl},\ }\bibfield  {journal} {\bibinfo
  {journal} {SciPost Physics Codebases}\ }\href
  {https://doi.org/10.21468/scipostphyscodeb.54} {10.21468/scipostphyscodeb.54}
  (\bibinfo {year} {2025}{\natexlab{a}})\BibitemShut {NoStop}%
\bibitem [{\citenamefont {Loizeau}\ \emph
  {et~al.}(2025{\natexlab{b}})\citenamefont {Loizeau}, \citenamefont {Buča},\
  and\ \citenamefont {Sels}}]{loizeau2025buca}%
  \BibitemOpen
  \bibfield  {author} {\bibinfo {author} {\bibfnamefont {N.}~\bibnamefont
  {Loizeau}}, \bibinfo {author} {\bibfnamefont {B.}~\bibnamefont {Buča}},\
  and\ \bibinfo {author} {\bibfnamefont {D.}~\bibnamefont {Sels}},\ }\href
  {https://arxiv.org/abs/2503.07403} {\bibinfo {title} {{Opening Krylov space
  to access all-time dynamics via dynamical symmetries}}} (\bibinfo {year}
  {2025}{\natexlab{b}}),\ \Eprint {https://arxiv.org/abs/2503.07403}
  {arXiv:2503.07403 [quant-ph]} \BibitemShut {NoStop}%
\bibitem [{\citenamefont {Füllgraf}\ \emph {et~al.}(2025)\citenamefont
  {Füllgraf}, \citenamefont {Wang}, \citenamefont {Steinigeweg},\ and\
  \citenamefont {Gemmer}}]{Fullgraf2025}%
  \BibitemOpen
  \bibfield  {author} {\bibinfo {author} {\bibfnamefont {M.}~\bibnamefont
  {Füllgraf}}, \bibinfo {author} {\bibfnamefont {J.}~\bibnamefont {Wang}},
  \bibinfo {author} {\bibfnamefont {R.}~\bibnamefont {Steinigeweg}},\ and\
  \bibinfo {author} {\bibfnamefont {J.}~\bibnamefont {Gemmer}},\ }\href
  {https://arxiv.org/abs/2503.17555} {\bibinfo {title} {{Lanczos-Pascal
  approach to correlation functions in chaotic quantum systems}}} (\bibinfo
  {year} {2025}),\ \Eprint {https://arxiv.org/abs/2503.17555} {arXiv:2503.17555
  [cond-mat.stat-mech]} \BibitemShut {NoStop}%
\bibitem [{\citenamefont {Gamayun}\ \emph {et~al.}(2025)\citenamefont
  {Gamayun}, \citenamefont {Mir}, \citenamefont {Lychkovskiy},\ and\
  \citenamefont {Ristivojevic}}]{Gamayun2025hvu}%
  \BibitemOpen
  \bibfield  {author} {\bibinfo {author} {\bibfnamefont {O.}~\bibnamefont
  {Gamayun}}, \bibinfo {author} {\bibfnamefont {M.~A.}\ \bibnamefont {Mir}},
  \bibinfo {author} {\bibfnamefont {O.}~\bibnamefont {Lychkovskiy}},\ and\
  \bibinfo {author} {\bibfnamefont {Z.}~\bibnamefont {Ristivojevic}},\
  }\bibfield  {title} {\bibinfo {title} {{Exactly solvable models for universal
  operator growth}},\ }\bibfield  {journal} {\bibinfo  {journal} {Journal of
  High Energy Physics}\ }\textbf {\bibinfo {volume} {2025}},\ \href
  {https://doi.org/10.1007/JHEP07(2025)256} {10.1007/JHEP07(2025)256} (\bibinfo
  {year} {2025})\BibitemShut {NoStop}%
\bibitem [{\citenamefont {Shirokov}\ \emph {et~al.}(2025)\citenamefont
  {Shirokov}, \citenamefont {Khrushchev}, \citenamefont {Uskov}, \citenamefont
  {Dudinets}, \citenamefont {Ermakov},\ and\ \citenamefont
  {Lychkovskiy}}]{Shirokov2025}%
  \BibitemOpen
  \bibfield  {author} {\bibinfo {author} {\bibfnamefont {I.}~\bibnamefont
  {Shirokov}}, \bibinfo {author} {\bibfnamefont {V.}~\bibnamefont
  {Khrushchev}}, \bibinfo {author} {\bibfnamefont {F.}~\bibnamefont {Uskov}},
  \bibinfo {author} {\bibfnamefont {I.}~\bibnamefont {Dudinets}}, \bibinfo
  {author} {\bibfnamefont {I.}~\bibnamefont {Ermakov}},\ and\ \bibinfo {author}
  {\bibfnamefont {O.}~\bibnamefont {Lychkovskiy}},\ }\href
  {https://arxiv.org/abs/2503.24362} {\bibinfo {title} {Quench dynamics via
  recursion method and dynamical quantum phase transitions}} (\bibinfo {year}
  {2025}),\ \Eprint {https://arxiv.org/abs/2503.24362} {arXiv:2503.24362
  [cond-mat.str-el]} \BibitemShut {NoStop}%
\bibitem [{\citenamefont {Pinna}\ \emph {et~al.}(2025)\citenamefont {Pinna},
  \citenamefont {Lunt},\ and\ \citenamefont {von Keyserlingk}}]{plv-25}%
  \BibitemOpen
  \bibfield  {author} {\bibinfo {author} {\bibfnamefont {G.}~\bibnamefont
  {Pinna}}, \bibinfo {author} {\bibfnamefont {O.}~\bibnamefont {Lunt}},\ and\
  \bibinfo {author} {\bibfnamefont {C.}~\bibnamefont {von Keyserlingk}},\
  }\href {https://arxiv.org/abs/2505.00089} {\bibinfo {title} {Approximation
  theory for green's functions via the lanczos algorithm}} (\bibinfo {year}
  {2025}),\ \Eprint {https://arxiv.org/abs/2505.00089} {arXiv:2505.00089
  [quant-ph]} \BibitemShut {NoStop}%
\bibitem [{\citenamefont {Lunt}\ \emph {et~al.}(2025)\citenamefont {Lunt},
  \citenamefont {Kriecherbauer}, \citenamefont {McLaughlin},\ and\
  \citenamefont {von Keyserlingk}}]{lkmv-25}%
  \BibitemOpen
  \bibfield  {author} {\bibinfo {author} {\bibfnamefont {O.}~\bibnamefont
  {Lunt}}, \bibinfo {author} {\bibfnamefont {T.}~\bibnamefont {Kriecherbauer}},
  \bibinfo {author} {\bibfnamefont {K.~T.-R.}\ \bibnamefont {McLaughlin}},\
  and\ \bibinfo {author} {\bibfnamefont {C.}~\bibnamefont {von Keyserlingk}},\
  }\href {https://arxiv.org/abs/2504.18311} {\bibinfo {title} {Emergent random
  matrix universality in quantum operator dynamics}} (\bibinfo {year} {2025}),\
  \Eprint {https://arxiv.org/abs/2504.18311} {arXiv:2504.18311 [quant-ph]}
  \BibitemShut {NoStop}%
\bibitem [{\citenamefont {Capizzi}\ \emph {et~al.}(2025)\citenamefont
  {Capizzi}, \citenamefont {Wang}, \citenamefont {Xu}, \citenamefont {Mazza},\
  and\ \citenamefont {Poletti}}]{cwxmp-25}%
  \BibitemOpen
  \bibfield  {author} {\bibinfo {author} {\bibfnamefont {L.}~\bibnamefont
  {Capizzi}}, \bibinfo {author} {\bibfnamefont {J.}~\bibnamefont {Wang}},
  \bibinfo {author} {\bibfnamefont {X.}~\bibnamefont {Xu}}, \bibinfo {author}
  {\bibfnamefont {L.}~\bibnamefont {Mazza}},\ and\ \bibinfo {author}
  {\bibfnamefont {D.}~\bibnamefont {Poletti}},\ }\bibfield  {title} {\bibinfo
  {title} {Hydrodynamics and the eigenstate thermalization hypothesis},\ }\href
  {https://doi.org/10.1103/PhysRevX.15.011059} {\bibfield  {journal} {\bibinfo
  {journal} {Phys. Rev. X}\ }\textbf {\bibinfo {volume} {15}},\ \bibinfo
  {pages} {011059} (\bibinfo {year} {2025})}\BibitemShut {NoStop}%
\bibitem [{\citenamefont {Su}\ \emph {et~al.}(1979)\citenamefont {Su},
  \citenamefont {Schrieffer},\ and\ \citenamefont {Heeger}}]{SSH-79}%
  \BibitemOpen
  \bibfield  {author} {\bibinfo {author} {\bibfnamefont {W.~P.}\ \bibnamefont
  {Su}}, \bibinfo {author} {\bibfnamefont {J.~R.}\ \bibnamefont {Schrieffer}},\
  and\ \bibinfo {author} {\bibfnamefont {A.~J.}\ \bibnamefont {Heeger}},\
  }\bibfield  {title} {\bibinfo {title} {Solitons in polyacetylene},\ }\href
  {https://doi.org/10.1103/PhysRevLett.42.1698} {\bibfield  {journal} {\bibinfo
   {journal} {Phys. Rev. Lett.}\ }\textbf {\bibinfo {volume} {42}},\ \bibinfo
  {pages} {1698} (\bibinfo {year} {1979})}\BibitemShut {NoStop}%
\bibitem [{\citenamefont {Su}\ \emph {et~al.}(1980)\citenamefont {Su},
  \citenamefont {Schrieffer},\ and\ \citenamefont {Heeger}}]{SSH-80}%
  \BibitemOpen
  \bibfield  {author} {\bibinfo {author} {\bibfnamefont {W.~P.}\ \bibnamefont
  {Su}}, \bibinfo {author} {\bibfnamefont {J.~R.}\ \bibnamefont {Schrieffer}},\
  and\ \bibinfo {author} {\bibfnamefont {A.~J.}\ \bibnamefont {Heeger}},\
  }\bibfield  {title} {\bibinfo {title} {Soliton excitations in
  polyacetylene},\ }\href {https://doi.org/10.1103/PhysRevB.22.2099} {\bibfield
   {journal} {\bibinfo  {journal} {Phys. Rev. B}\ }\textbf {\bibinfo {volume}
  {22}},\ \bibinfo {pages} {2099} (\bibinfo {year} {1980})}\BibitemShut
  {NoStop}%
\end{thebibliography}%
\end{document}